\documentclass[a4paper]{article}
\usepackage[a4paper,top=3cm,bottom=2cm,left=3cm,right=3cm,marginparwidth=1.75cm]{geometry}
\usepackage{amsfonts}
\usepackage{bm}
\usepackage{extarrows}
\usepackage{amssymb}
\usepackage{color}
\usepackage[all]{xy}
\usepackage{graphicx}
\usepackage{braket}
\usepackage{amsmath}
\usepackage{appendix}
\usepackage{graphicx}
\usepackage[colorinlistoftodos]{todonotes}
\usepackage{amsthm}
\usepackage{mathrsfs}
\usepackage{amssymb}
\usepackage{accents}
\usepackage{extarrows}
\usepackage{makecell}
\usepackage{authblk}
\usepackage{url}
\usepackage{hyperref}
\usepackage{cite}
\usepackage{eufrak}
\hypersetup{colorlinks,
            citecolor=black,
            linkcolor=black,
            urlcolor=green,
            pdftex}

\usepackage[english]{babel}
\usepackage[utf8x]{inputenc}
\usepackage[T1]{fontenc}

\usepackage[normalem]{ulem}

\title{Quantum Geometry Effects in Quantum Field Theory: Hamiltonian constraint Generates Gravity-Matter Entanglement}
\author[1]{Gaoping Long }

\author[2]{Cong Zhang \footnote{corresponding author:  cong.zhang@bnu.edu.cn}}

\affil[1]{College of Physics $\&$ Optoelectronic Engineering, Jinan University, Guangzhou, 510632, Guangdong, China}
\affil[2]{School of Physics and Astronomy, Key Laboratory of Multiscale Spin Physics,
Ministry of Education, Beijing Normal University, Beijing 100875, China}

\date{}

\begin{document}

\maketitle

\begin{abstract}

In this paper, we address a foundational challenge in quantum field theory on curved spacetime by developing a consistent framework within loop quantum gravity. We introduce a methodology for defining meaningful superpositions of quantum geometry and matter states. This is achieved by identifying a restricted subspace of the gravitational phase space, which ensures unitary equivalence among Fock representations of a scalar field across different quantum geometries. Within the resulting well-defined state space, we derive weak solutions to the quantum Hamiltonian constraint of general relativity.  Furthermore, we generalize the Hartle-Hawking vacuum state to this quantum geometric framework. The resulting state exhibits the inherent entanglement between geometry and matter, which arises from the quantum Hamiltonian constraint of general relativity.
This work establishes a principled framework for studying geometry-matter entanglement and offers new insights into the quantum foundations of the black hole information paradox.
\end{abstract}
\section{Introduction}

The black hole information paradox remains one of the most profound challenges at the interface of quantum mechanics (QM) and general relativity (GR). Specifically, while black holes are among the most significant predictions of GR, Hawking’s discovery of thermal radiation from evaporating black holes suggests that information about the initial state of matter may be lost during evaporation—thereby violating the unitarity of QM \cite{Hawking:1974rv,Hawking:1975vcx,Parikh:1999mf,Parikh:2004ih,Hawking:1976ra,Unruh:2017uaw,Mathur:2009hf,Almheiri:2020cfm}.    A central development in addressing this paradox is the derivation of the Page curve, which describes the time evolution of the entanglement entropy of Hawking radiation under the assumption of unitary evaporation \cite{Page:1993wv,Page:2013dx,Penington:2019npb}. The Page curve implies that, at late times, the entropy of the radiation is suppressed as it becomes purified by correlations with the remaining black hole interior—a process that necessarily requires entanglement between the radiation field and the underlying spacetime geometry.
This insight strongly suggests that a complete resolution of the information paradox must account for the quantum nature of geometry itself. In a full quantum gravity framework, the total state of the system should take the form of an entangled superposition
\begin{eqnarray}\label{entangle111}
 | \Psi\rangle = \sum_{\mathfrak{g},\phi} c_{\mathfrak{g},\phi}|\mathfrak{g}\rangle\otimes |\phi\rangle, 
 \end{eqnarray} 
 where $|\mathfrak{g}\rangle$ denotes a quantum state of spacetime geometry and $|\phi\rangle$ a state of the matter field. However, while the Page curve has been successfully reproduced using semiclassical gravitational path integrals and replica wormhole techniques \cite{Almheiri:2020cfm,Almheiri:2019qdq,Penington:2019npb,Almheiri:2019hni}, these approaches do not yet provide a dynamical unitary evolution of a state like \eqref{entangle111} within a concrete Hilbert space of quantum gravity. More fundamentally, the very definition of the product states $|\mathfrak{g}\rangle\otimes |\phi\rangle$ remains ambiguous: what does it mean to define a quantum field on a quantum geometry?

Historically, quantum field theory on curved spacetime (QFTCS) sidesteps this issue by treating geometry as a fixed classical background $\mathfrak{g}$, on which matter states  $|\phi\rangle_\mathfrak{g}$ are constructed \cite{Hollands:2014eia,Wald:1995yp,Birrell_Davies_1982}. This approach is perturbative in nature, since the backreaction of the energy-momentum tensor $\hat{T}_{\mu\nu}$ of the quantum matter field on the classical geometry is ignored or effectively treated via the effective Einstein equation $G_{\mu\nu}=8\pi G \langle \hat{T}_{\mu\nu}\rangle$. However, from the perspective of quantum gravity, distinct matter states generically correspond to distinct geometric configurations, even if their classical limits coincide. In particular, these microscopic geometric distinctions may encode the microstates responsible for black hole entropy \cite{Bekenstein:1973ur,Bekenstein:1974ax,Wald:1999vt,Ashtekar:1997yu,Carlip:2008wv,Harlow:2014yka}. Consequently, when considering the entanglement between geometry and matter, the QFTCS prescription—building all $|\phi\rangle$ on a single classical $\mathfrak{g}$—becomes ill-defined, as it conflates physically distinct quantum geometries.

To construct the entangled state \eqref{entangle111} rigorously, three interrelated obstacles must be overcome:
\begin{enumerate}
    \item A well-defined Hilbert space $\mathcal{H}_{\text{g}}$ for quantum geometry, containing states $|\mathfrak{g}\rangle$ that admit a semiclassical interpretation;
    \item A consistent formulation of the matter Hilbert space $\mathcal{H}^\phi$ on each quantum geometry $|\mathfrak{g}\rangle$, such that the superpositions of the product states $|\mathfrak{g}\rangle\otimes |\phi\rangle$ across different $|\mathfrak{g}\rangle$ are mathematically meaningful;
    \item The product states $|\mathfrak{g}\rangle\otimes |\phi\rangle$ must satisfy the quantum constraints of GR, reflecting its nature as a totally constrained Hamiltonian system \cite{Arnowitt:1962hi,Arnowitt:1960es }.
\end{enumerate}

In this work, we address these challenges within the framework of loop quantum gravity (LQG)—a non-perturbative, background-independent approach to quantum gravity that provides a rigorously defined kinematical Hilbert space for geometry \cite{Ashtekar2012Background,Han2005FUNDAMENTAL,first30years,Long0propagator,Zhang:2022vsl,Longrepresentationtwisted,Bianchi:2008es,Ma:2010fy,Long_2025,Giesel_2006Consistencycheck,Yang_2019Consistencycheck,Long:2021izw,QoperatorPhysRevD.62.104021,PhysRevD.103.086016,Han:2024ydv,volumePhysRevD.94.044003,long2020operators}. LQG has already yielded microscopic derivations of black hole entropy through counting of boundary states \cite{Ashtekar:1997yu,Ashtekar:2000eq, Long:2024lbd,Song:2020arr,Kaul:2000kf,Ghosh:2011fc,Basu:2009cw,Engle:2010kt,Song:2022zit,Perez:2014ura}, and its polymer quantization scheme resolves the classical singularity in symmetry-reduced models, yielding bouncing cosmology models and regular black hole interiors \cite{Ashtekar:2011ni,Long:2020oma,Zhang:2021zfp,Bojowald:2001xe, Ashtekar:2006wn,Ashtekar:2005qt, Gambini:2013hna,Corichi:2015xia,Dadhich:2015ora,Olmedo:2017lvt,Ashtekar:2018lag,Han:2020uhb,Kelly:2020lec,Han:2022rsx,Giesel:2023hys,Han:2024rqb,PhysRevLett.102.051301}.
Specifically, we work in the spherically symmetric model of gravity (GSSM) and employ semiclassical coherent states $|\mathfrak{g}\rangle$ in loop quantum theory for gravitational quantum states, with $|\mathfrak{g}\rangle$ being sharply peaked at classical phase space points and satisfying the Ehrenfest property \cite{Long:2024lbd,Long:2020euh,PhysRevD.104.046014,Calcinari_2020,
Thiemann:2000bx,
long2019coherent,Zhang:2021qul,
Han:2019vpw, Long:2021lmd,Long:2022cex}. To tackle the second obstacle, we recognize that the Fock representation of a scalar field depends on the background geometry. For arbitrary $|\mathfrak{g}\rangle$ and $|\mathfrak{g}'\rangle$, the corresponding Fock spaces $\mathcal{H}^\phi_{\mathfrak{g}}$ and $\mathcal{H}^\phi_{\mathfrak{g}'}$ are generally unitarily inequivalent, rendering superpositions like \eqref{entangle111} ill-defined \cite{Wald:1995yp}. We resolve this by identifying a restricted subspace $\mathcal{U}_{v,v'}$ of the gravitational phase space, in which the Bogoliubov transformation between Fock spaces $\mathcal{H}^\phi_{\mathfrak{g}}$ and $\mathcal{H}^\phi_{\mathfrak{g}'}$  satisfies the Shale–Stinespring criterion (i.e., the $\beta$-coefficients form a Hilbert–Schmidt operator) \cite{Shale1962,Wald:1995yp}. In this restricted subset $\mathcal{U}_{v,v'}$, a common Fock space $\mathcal{H}^\phi$ exists, enabling a quantum state space $\tilde{\mathcal{H}}_{\text{g}}\otimes\mathcal{H}^\phi$ in which superpositions like \eqref{entangle111} are well-defined.
The third obstacle stems from the fact that GR, in its Hamiltonian formulation, is governed entirely by the ADM constraints--the diffeomorphism and Hamiltonian constraints \cite{Arnowitt:1962hi,Arnowitt:1960es }. Hence, physical states must lie in the kernel of quantum constraint operators. Rather than seeking strong solutions (which remain elusive even in reduced models), we adopt a weak constraint approach. In  detail, we require that matrix elements of the quantum Hamiltonian constraint vanish between selected coherent–Fock states. This yields a family of approximate physical states that are inherently entangled between quantum geometries and matter.

Our results show that weak solutions to the quantum Hamiltonian constraint, which are constructed from gravitational coherent states $|\mathfrak{g}\rangle$ and scalar Fock states $|\phi\rangle$, naturally encode entanglement between geometry and matter. Specifically, when the coherent state $|\mathfrak{g}\rangle$ is peaked at a semiclassical geometry containing an apparent horizon, the admissible Fock modes (defined in a modified tortoise coordinate) are correlated with the fluctuating wave packets on the gravitational phase space. Crucially, this correlation is not imposed by hand, but emerges as a direct consequence of the imposition of the constraint of GR. 
Moreover, we consider a generalized Hartle–Hawking vacuum on quantum geometry by superposing the weak solutions of the quantum Hamiltonian constraint, which takes precisely the entangled form \eqref{entangle111}. 
Generally, this work  provides a concrete realization of how the constraints of GR enforce geometry–matter entanglement, and it may offer a first-principles foundation for understanding the quantum origin of the Page curve and the unitary evaporation of black holes.

This paper is organized as follows. In  Sec. \ref{sec201}, we will review the classical theory of the gravitationally spherically symmetric  model coupled to a massless scalar field in its Hamiltonian formulation;
Then,  in Sec. \ref{sec202},  we will consider the quantum scalar field on a classical geometry and highlight a fundamental issue inherent in quantum field theory (QFT)  on a fixed classical background geometry.  To tackle this issue, we will turn to quantum gravity in Sec. \ref{sec3}.  Specifically, in Sec. \ref{sec301}, we will discuss the unitary equivalence among Fock representations for the scalar field across different quantum geometries; Then, the well-defined weak solution space of the quantum Hamiltonian constraint will be given in Sec. \ref{sec302}; 
  Further, in Sec. \ref{sec303},  we will establish a generalized Hartle–Hawking vacuum on quantum geometry by superposing weak solutions of the quantum Hamiltonian constraint and analyze its properties.
  Finally, we conclude in Sec. \ref{sec4} with a discussion of our results and future directions.

\section{Gravitational spherical symmetric  model coupling to massless scalar field }\label{sec2}
\subsection{Classical theory}\label{sec201}
 The Hamiltonian formulation of the spherically symmetric GR can be formulated as a connection theory. Specifically, the spatial 3-manifold is denoted as $\Sigma=\mathbb X\times\mathbb S^2$ carrying the $SU(2)$ connection field, where $\mathbb X$ denotes a 1-D manifold and $\mathbb S^2$ is the 2-sphere.  Let $(x,\theta,\varphi)$ denote the coordinates of $\Sigma$ adapted to the  $SU(2)$ connection field.  The phase space of the gravitational field, denoted by $\mathcal P^{\text{g}} $, comprises of fields $(K_x,E^x,K_\varphi,E^\varphi)$ defined on the quotient manifold $\Sigma/ SU(2) \cong \mathbb X$ (see, e.g., \cite{Bojowald:2005cb,Gambini2023} for more details of the kinematical structure). The Ashtekar-Barbero variables $(A_a^i,E^b_j)$ (see, e.g., \cite{Zhang:2024khj,Zhang:2024ney}) read
\begin{equation}\label{eq:symmetricAE}
\begin{aligned}
A_a^i\tau_i \text{d}x^a&=K_a^i\tau_i\text{d}x^a+\Gamma_a^i\tau_i\text{d}x^a,\\
E^a_i\tau^i\partial_a&=E^x\sin\theta\tau_3\partial_x+E^\varphi\sin\theta\tau_1\partial_\theta+E^\varphi\tau_2\partial_\varphi,
\end{aligned}
\end{equation}
where $\Gamma_a^i$ is the spin connection compatible with the densitized triad $E^a_i$, and $K_a^i$ is the extrinsic curvature 1-form given by
\begin{equation}\label{eq:symmetricK}
K_a^i\tau_i\text{d}x^a=K_x\tau_3\text{d}x+K_\varphi\tau_1\text{d}\theta+K_\varphi\sin\theta\tau_2\text{d}\varphi.
\end{equation}
Here we use the convention $\tau_j=-i\sigma_j/2$ with $\sigma_j$ being the Pauli matrices. The spatial metric on $\Sigma$ is related to the variables $(A_a^i,E^b_j)$ by
\begin{equation}
ds^2=\frac{(E^\varphi)^2}{|E^x|}dx^2+|E^x| (d\theta^2+\sin^2\theta d\varphi^2).
\end{equation}
The non-vanishing Poisson brackets between the phase space variables are
\begin{equation}
\begin{aligned}
\{K_x(x),E^x(x')\}=&2G\delta(x-x'),\\
\{K_\varphi(x),E^\varphi(x')\}=&G\delta(x-x'),
\end{aligned}
\end{equation}
where $G$ is the gravitational constant.

The Hamiltonian formulation of a massless scalar field in the spherical symmetric  model is based on the phase space $\mathcal P^{\phi} $, which is composed by the fields $(\phi,\pi_\phi)$ defined on the quotient manifold $ \mathbb X$ , and equipped with the non-vanishing Poisson brackets,
\begin{equation}\label{Poi1110}
\{\phi(x),\pi_\phi(x')\}=\delta(x-x').
\end{equation}
Then, the total phase space of the spherically symmetric  model of gravity coupled to the massless scalar field is given as $\mathcal P^{\text{g}} \times \mathcal P^{\phi}$.  Moreover, this system is totally constrained and is governed by two sets of constraints: the diffeomorphism constraint $H_x$ and the Hamiltonian constraint $H$. They are expressed as 
\begin{equation}
H_x=H_x^{\text{g}}+{H}^\phi_x\approx0
\end{equation}
and 
\begin{equation}
H=H^{\text{g}}+H^{\phi}\approx0
\end{equation}
respectively, where   \cite{Zhang:2024khj,Zhang:2024ney}
\begin{equation}
H^{\text{g}}=-2\frac{E^\varphi}{\partial_xE^x}\partial_x M+2\frac{\sqrt{E^x}K_\varphi}{\partial_x E^x}H^{\text{g}}_x, \quad  H^{\text{g}}_x=\frac{1}{2G}(-K_x\partial_x E^x+2E^\varphi\partial_x K_\varphi)
\end{equation}
with $M=\frac{\sqrt{E^x}}{2G}(1+(K_\varphi)^2-\frac{(\partial_x E^x)^2}{4(E^\varphi)^2})$, and

\begin{equation}
H^{\phi}=\frac{1}{2}(\frac{\pi_{\phi}^2}{\sqrt{E^x}E^\varphi}+\frac{(E^x)^{3/2}}{E^\varphi}(\partial_x\phi)^2),\quad {H}^\phi_x=\pi_\phi\partial_x\phi.
\end{equation}

To obtain the quantum physical states, one needs to quantize the phase space  $\mathcal P^{\text{g}} \times \mathcal P^{\phi}$ to get the corresponding Hilbert space $\mathcal{H}_{\text{g}}\otimes \mathcal{H}^{\phi}$,  and then solve the quantum versions of the diffeomorphism constraint $H_x$ and the Hamiltonian constraint $H$. In this paper, we  focus only on the  Hamiltonian constraint $H$, which can be reformulated as
\begin{eqnarray}
H&=&-2\frac{E^\varphi}{\partial_xE^x}\partial_x M+2\frac{\sqrt{E^x}K_\varphi}{\partial_x E^x}H^{\text{g}}_x+H^\phi\\\nonumber
&=&-2\frac{E^\varphi}{\partial_xE^x}\left(\partial_x M-\frac{\sqrt{E^x}}{ E^\varphi}K_\varphi H^{\text{g}}_x-\frac{\partial_xE^x}{2E^\varphi}H^\phi\right)
\\\nonumber
&=:&-2\frac{E^\varphi}{\partial_xE^x}\tilde{H}.
\end{eqnarray}
To carry out a further discussion, we consider the smeared version of $\tilde{H}$ , which reads
\begin{equation}
\tilde{H}^\delta(v,v'):=\int^{x(v)+\delta}_{x(v)} dx \tilde{H}=( M_{v'}-M_{v})-\tilde{H}^\delta_{\phi}(v,v'),
\end{equation}
where $v$ is an arbitrary point on the 1-dimensional manifold $\mathbb X$ , $v'$ is another point that satisfies $x(v')-x(v)=\delta$ with $\delta>0$ being a finite small real number, and $\tilde{H}^\delta_{\phi}(v,v')$ is defined by
\begin{eqnarray}
\tilde{H}^\delta_{\phi}(v,v')&:=&\int^{x(v)+\delta}_{x(v)}  dx(\frac{\sqrt{E^x}}{ E^\varphi}K_\varphi H^{\text{g}}_x+\frac{\partial_xE^x}{2E^\varphi}H^\phi).
\end{eqnarray}
It is easy to see that the Hamiltonian constraint equations $H=0$ are solved if and only if the equations 
\begin{equation}\label{cons111}
\tilde{H}^\delta(v,v')=( M_{v'}-M_{v})-\tilde{H}^\delta_{\phi}(v,v')=0
\end{equation} 
are solved for all $v\in \Sigma$ and all $\delta\in \mathbb{R}$. In the following part of this article, we will focus on the solution of the quantum version of Eq.\eqref{cons111}.

\subsection{Issue of quantum field theory on classical geometry}\label{sec202}

The quantum theory of the scalar field on a specific classical spacetime geometry can be established by effectively solving Eq.\eqref{cons111}.  In other words, one can fix the geometry components and promote the scalar field component as the quantum counterpart in Eq.\eqref{cons111}, which leads to the effective Hamiltonian constraint equation
\begin{eqnarray}\label{cons444}
{ M}_{v'}(\mathfrak{g})-{ M}_{v}(\mathfrak{g})=\frac{\langle\phi'|\widehat{\tilde{H}^\delta_{\phi}}(\mathfrak{g};v,v')|\phi\rangle}{\langle\phi'|\phi\rangle}
\end{eqnarray}
where  $\mathfrak{g}=\{E^x(\mathfrak{g}),K_x(\mathfrak{g}),E^\varphi(\mathfrak{g}),E_\varphi(\mathfrak{g})\}$ is a phase space point in $\mathcal{P}^{\text{g}}$,  ${ M}_{v'}(\mathfrak{g}), { M}_{v}(\mathfrak{g})$ is the values of ${ M}_{v'}, { M}_{v}$ at the phase space point labeled by $\mathfrak{g}\in \mathcal{P}^{\text{g}}$, and $\widehat{\tilde{H}^\delta_{\phi}}(\mathfrak{g};v,v')$ is defined by
\begin{eqnarray}\label{Hgvv}
\widehat{\tilde{H}^\delta_{\phi}}(\mathfrak{g};v,v')=\int^{x(v)+\delta}_{x(v)}  dx\left(\left(\left.\frac{\sqrt{E^x}}{ E^\varphi}K_\varphi\right|_{\mathfrak{g}}\right) {H}^{\text{g}}_x(\mathfrak{g})+\left(\left.\frac{\partial_xE^x}{2E^\varphi}\right|_{\mathfrak{g}}\right)\hat{H}^\phi(\mathfrak{g})\right)
\end{eqnarray}
with ${H}^{\text{g}}_x(\mathfrak{g})$ being the value of  ${H}^{\text{g}}_x$ at the phase space point $\text{g}$,
and $\hat{H}^{\phi}(\mathfrak{g})$ being defined by
\begin{eqnarray}\label{scalarphig}
\hat{H}^{\phi}(\mathfrak{g}):=\frac{1}{2}\left(    
    \left(\left.  \frac{1}{\sqrt{E^x}E^\varphi}    \right|_{\mathfrak{g}}\right)(\hat{\pi}_{\phi})^2+         \left(\left.  \frac{(E^x)^{3/2}}{E^\varphi}    \right|_{\mathfrak{g}}\right)   (\partial_x\hat{\phi})^2\right).
\end{eqnarray}
Then, for the given spatial (intrinsic and extrinsic) geometry $\mathfrak{g}$, the quantum scalar field in the region $x(v)<x< x(v)+\delta$ must satisfy Eq.\eqref{cons444}. 
Generally, it is too complicated to solve Eq.\eqref{cons444} for some arbitrary  $\mathfrak{g}, v,v'$, and thus we consider the solution for some special  $\mathfrak{g}$.
Let us first fix the gauge by setting
\begin{equation}\label{gauge1}
E^x=x^2
\end{equation}
 for all $\mathfrak{g}$, and solve the diffeomorphism constraint $H_x=0$ which leads to
\begin{equation}\label{gauge2}
K_x=\frac{2E^\varphi\partial_x K_\varphi+{2G}\pi_\phi\partial_x\phi}{\partial_x E^x}.
\end{equation}
Now, the remaining dynamical variables  contain only the pairs $\mathfrak{g}=(K_\varphi,E^\varphi)$ and $(\phi,\pi_\phi)$. From now on, let us focus on the special collection
\begin{equation}
\mathcal{S}_{ v,v'}:=\{\mathfrak{g}\in \mathcal{P}^{\text{g}}|\  \mathfrak{g} \ \text{satisfying\  the\ following\  conditions\ (1) and   (2);}\}
\end{equation}
\begin{enumerate}\label{condition333}
\item  The spatial (intrinsic and extrinsic) geometry  $\mathfrak{g}$  ensures an apparent horizon at  $x=x_{\text{h}}=x(v)$; In other words, the expansion $(\frac{\partial_x |E^x|}{\sqrt{|E^x|}E^\varphi}-\frac{2K_\varphi}{\sqrt{|E^x|}})|_{\mathfrak{g}}$ for $\mathfrak{g}$ vanishes at  $x=x_{\text{h}}$;
\item   In the gauge that $K_\varphi=0$ and $E^\varphi>0$  in the region $x_{\text{h}}\leq x\leq x_{\text{h}}+\delta$, the spatial (intrinsic and extrinsic) geometry  $\mathfrak{g}$ satisfies
\begin{equation}
\int_{x_{\text{h}}}^{x_{\text{h}}+\delta} dx \frac{(E^\varphi)^2}{x^2}\to \infty.
\end{equation}
\end{enumerate}
From now on, let us fix the gauge  $K_\varphi=0$ and $E^\varphi>0$  in the region $x_{\text{h}}\leq x\leq x_{\text{h}}+\delta$, and then each $\mathfrak{g}\in\mathcal{S}_{ v,v'}$ is totally determined by $E^\varphi$  in $x_{\text{h}}\leq x\leq x_{\text{h}}+\delta$. Especially,  by introducing  the notation  $h_{\mathfrak{g}}(x):=\left.(E^\varphi)^2\right|_{\mathfrak{g}}$ , one has $\frac{x^2}{h_{\mathfrak{g}}(x)}\to0$ for  $x\to x_{\text{h}}$.

In order to represent the operator $\hat{H}^{\phi}(\mathfrak{g})$   and find a state $|\phi\rangle$ solving Eq.\eqref{cons444} for $\mathfrak{g}\in\mathcal{S}_{ v,v'}$. Let us consider the quantum theory of the scalar field in the Fock representation.
Especially, one should notice that the quantization of the scalar field in the Fock representation relies on a choice of the coordinate system. Indeed, one can choose an arbitrary coordinate system to construct the Fock space, and the Fock spaces associated with different coordinate systems can be   related by the Bogoliubov transformation. As we show in the following construction, the solutions of Eq.\eqref{cons444} can be found in the Fock spaces associated to a modified tortoise coordinate associated with $\mathfrak{g}$. 

For an arbitrary spatial (intrinsic and extrinsic) geometry  $\mathfrak{g}\in \mathcal{S}_{ v,v'}$, we can introduce the tortoise coordinate $x^\ast_{\mathfrak{g}}$ by the transformation  
\begin{equation}
\frac{\partial x}{\partial {x^\ast_{\mathfrak{g}}}}=\frac{x^2}{h_{\mathfrak{g}}(x)},
\end{equation}
 with  the tortoise coordinate $x^\ast_{\mathfrak{g}}$   satisfying
\begin{equation}
\lim_{x\to x_{\text{h}}} x^\ast_{\mathfrak{g}}\to-\infty, \quad \lim_{x\to x_{\text{h}}+\delta} x^\ast_{\mathfrak{g}}\to\delta^\ast_{\mathfrak{g}},\quad 0<\delta^\ast_{\mathfrak{g}}<+\infty.
\end{equation}

In the tortoise coordinate $x^\ast_{\mathfrak{g}}$, the spatial metric in the region $x_{\text{h}}< x\leq x_{\text{h}}+\delta$ can be written as
\begin{equation}
ds^2=\frac{x^2}{h_{\mathfrak{g}}(x)}(dx^\ast_{\mathfrak{g}})^2+x^2 (d\theta^2+\sin^2\theta d\varphi^2).
\end{equation}
Now,  the non-vanishing quantum commutator corresponding to \eqref{Poi1110} 
in the tortoise coordinate $x^\ast_{\mathfrak{g}}$ is given by  
\begin{equation}\label{quan111}
[\hat{\phi}(x^\ast_{\mathfrak{g}}),\hat{\pi}_\phi(x'^\ast_{\mathfrak{g}})]=\mathbf{i}\hbar\delta(x^\ast_{\mathfrak{g}}-x'^\ast_{\mathfrak{g}}).
\end{equation}
One can  define the annihilation operator $a^{\mathfrak{g}}_k$ as
\begin{equation}\label{ak}
a^{\mathfrak{g}}_k:=\lim_{L\to\infty}\int_{-L}^{+L} d\check{x}^\ast_{\mathfrak{g}} {e^{-\mathbf{i}k\check{x}^\ast_{\mathfrak{g}}}}\frac{1}{\sqrt{2\omega_k}}({x}\cdot\omega_k\hat{\phi}(\check{x}^\ast_{\mathfrak{g}})+\mathbf{i}  \frac{\hat{\pi}_\phi(\check{x}^\ast_{\mathfrak{g}})}{{x}})
\end{equation}
where $\omega_k:=|k|$ , $x=x(\check{x}^\ast_{\mathfrak{g}})$ and 
\begin{equation}\label{modified}
\check{x}^\ast_{\mathfrak{g}}=x^\ast_{\mathfrak{g}}+L-\delta^\ast_{\mathfrak{g}}.
\end{equation}
Correspondingly, the Hermitian conjugate gives the creation operator $a^{\mathfrak{g},\dagger}_k$ as
\begin{equation}\label{ak+}
a^{\mathfrak{g},\dagger}_k:=\lim_{L\to\infty}\int_{-L}^{+L}  d\check{x}^\ast_{\mathfrak{g}}{e^{\mathbf{i}k\check{x}^\ast_{\mathfrak{g}}}}\frac{1}{\sqrt{2\omega_k}}({x}\cdot\omega_k\hat{\phi}(\check{x}^\ast_{\mathfrak{g}})-\mathbf{i}  \frac{\hat{\pi}_\phi(\check{x}^\ast_{\mathfrak{g}})}{{x}}).
\end{equation}
Then, the operators $\hat{\phi}(\check{x}^\ast_{\mathfrak{g}})$ and $\hat{\pi}_\phi(\check{x}^\ast_{\mathfrak{g}})$  in the region $-\infty <x^\ast_{\mathfrak{g}} <\delta^\ast_{\mathfrak{g}}$ can be given as
\begin{equation}\label{phiaa}
\hat{\phi}(\check{x}^\ast_{\mathfrak{g}})\overset{L\to\infty}{=} \int \frac{dk}{2\pi}\frac{1}{\sqrt{2\omega_k}} (\frac{e^{\mathbf{i}k\check{x}^\ast_{\mathfrak{g}}}}{x}a^{\mathfrak{g}}_k+\frac{e^{-\mathbf{i}k\check{x}^\ast_{\mathfrak{g}}}}{x}a^{\mathfrak{g},\dagger}_k)
\end{equation}
and 
\begin{equation}\label{piphiaa}
\hat{\pi}_\phi(\check{x}^\ast_{\mathfrak{g}})\overset{L\to\infty}{=} (-\mathbf{i})\int  \frac{dk}{2\pi}\sqrt{\frac{\omega_k}{2}} {x} ({e^{\mathbf{i}k\check{x}^\ast_{\mathfrak{g}}}}a^{\mathfrak{g}}_k-{e^{-\mathbf{i}k\check{x}^\ast_{\mathfrak{g}}}}a^{\mathfrak{g},\dagger}_k)
\end{equation}
where we used 
\begin{equation}
\lim_{L\to\infty}\int_{-L}^{+L}  d\check{x}^\ast_{\mathfrak{g}}{e^{\mathbf{i}k\check{x}^\ast_{\mathfrak{g}}}}
=2\pi \delta(k).
\end{equation}
Furthermore, by using \eqref{quan111}, one can check that
\begin{equation}
[a^{\mathfrak{g}}_k,a^{\mathfrak{g},\dagger}_{k'}]\overset{L\to\infty}{=} 2\pi \hbar\delta(k-k'), \quad [a^{\mathfrak{g}}_k,a^{\mathfrak{g}}_{k'}]\overset{L\to\infty}{=} [a^{\mathfrak{g},\dagger}_k,a^{\mathfrak{g},\dagger}_{k'}]\overset{L\to\infty}{=} 0.
\end{equation}
With this algebra between $a^{\mathfrak{g}}_k$ and $a^{\mathfrak{g},\dagger}_{k'}$   established, one can define the  vacuum state (Boulware-like) $|0_{\text{B}}\rangle_{\mathfrak{g}}$ in the region $-\infty <{x}_\ast <\delta^\ast_{\mathfrak{g}}$ by 
\begin{equation}
a^{\mathfrak{g}}_k|0_{\text{B}}\rangle_{\mathfrak{g}}=0,\quad \forall k.
\end{equation}
Correspondingly, the one-particle state can be defined as
\begin{equation}
|k\rangle_{\mathfrak{g}}:=a^{\mathfrak{g},\dagger}_k|0_{\text{B}}\rangle_{\mathfrak{g}}
\end{equation}
with the inner product
\begin{equation}
\langle k'|k\rangle_{\mathfrak{g}}=\delta(k'-k).
\end{equation}
Then, one can introduce the Hilbert space $\mathcal{H}^\phi_{\mathfrak{g}}$ of the field $\phi$ spanned by acting on the vacuum with all possible combinations of $a^{\mathfrak{g},\dagger}_k$, which reads
\begin{equation}
\mathcal{H}_{\mathfrak{g}}^\phi:=  \{|0_{\text{B}}\rangle_{\mathfrak{g}}, a^{\mathfrak{g},\dagger}_k|0_{\text{B}}\rangle_{\mathfrak{g}}, a^{\mathfrak{g},\dagger}_ka^{\mathfrak{g},\dagger}_{k'}|0_{\text{B}}\rangle_{\mathfrak{g}}, a^{\mathfrak{g},\dagger}_ka^{\mathfrak{g},\dagger}_{k'}a^{\mathfrak{g},\dagger}_{k''}|0_{\text{B}}\rangle_{\mathfrak{g}},...\}.
\end{equation}

It is worth to have some discussions on the construction of the Fock space $\mathcal{H}_{\mathfrak{g}}^\phi$. First, Eq.\eqref{modified} can be regarded as a coordinate transformation introducing the modified tortoise coordinate $\check{x}^\ast_{\mathfrak{g}}$. This modified tortoise coordinate $\check{x}^\ast_{\mathfrak{g}}$  ensures that the region $-\infty <x^\ast_{\mathfrak{g}} <\delta^\ast_{\mathfrak{g}}$ can be represented as $-\infty <\check{x}^\ast_{\mathfrak{g}} <+\infty$ in the limit $L\to\infty$. Hence, the Fock space $\mathcal{H}^\phi_{\mathfrak{g}}$  can be constructed in  the region $-\infty <x^\ast_{\mathfrak{g}}  <\delta^\ast_{\mathfrak{g}}$  based on the mode $e^{\mathbf{i}k\check{x}^\ast_{\mathfrak{g}}}$. 
Second, notice that the construction of  $\mathcal{H}_{\mathfrak{g}}^\phi$ is valid for all of $\mathfrak{g}\in \mathcal{S}_{v,v'}$. However, the Fock spaces $\mathcal{H}_{\mathfrak{g}}^\phi$ constructed on different $\mathfrak{g}$ may  not be  equivalent; In other words, one cannot ensure  unitary equivalence   $\mathcal{H}_{\mathfrak{g}}^\phi\cong \mathcal{H}_{\mathfrak{g}'}^\phi$ for arbitrary $\mathfrak{g}\neq \mathfrak{g}'$.

Now, it is ready to represent the right-hand side of Eq.\eqref{cons444} in   Hilbert space $\mathcal{H}^\phi_{\mathfrak{g}}$.  We first express the above construction in the coordinate $x$. Notice that  ${\pi}_\phi$ is a field with density one, and thus 
\begin{equation}
\hat{\pi}_\phi(x)=\frac{\partial x^\ast_{\mathfrak{g}}}{\partial {x}}\hat{\pi}_\phi(x^\ast_{\mathfrak{g}})=\frac{\partial x^\ast_{\mathfrak{g}}}{\partial {x}}\frac{\partial \check{x}^\ast_{\mathfrak{g}}}{\partial {x^\ast_{\mathfrak{g}}}}\hat{\pi}_\phi(\check{x}^\ast_{\mathfrak{g}}),
\end{equation}
which leads to
\begin{equation}
\hat{\pi}_\phi(x)=\frac{\partial \check{x}^\ast_{\mathfrak{g}}}{\partial {x}}\hat{\pi}_\phi(\check{x}^\ast_{\mathfrak{g}})=(-\mathbf{i}){h_{\mathfrak{g}}(x)}\int  \frac{dk}{2\pi}\sqrt{\frac{\omega_k}{2}}  (\frac{e^{\mathbf{i}k\check{x}^\ast_{\mathfrak{g}}}}{x}a^{\mathfrak{g}}_k-\frac{e^{-\mathbf{i}k\check{x}^\ast_{\mathfrak{g}}}}{x}a^{\mathfrak{g},\dagger}_k).
\end{equation}
Then, the Hamiltonian operator of $\phi$ is given by     
\begin{eqnarray}
\int_{x_{\text{h}}}^{x_{\text{h}}+\delta}  dx \left(\left.\frac{\partial_xE^x}{2E^\varphi}\right|_{\mathfrak{g}}\right)\hat{H}^\phi(\mathfrak{g})
&=&\int_{x_{\text{h}}}^{x_{\text{h}}+\delta} dx\frac{1}{2}(\frac{(\hat{\pi}_{\phi}(x))^2}{ {h_{\mathfrak{g}}(x)}}+{h_{\mathfrak{g}}(x)}(\partial_{\check{x}^\ast_{\mathfrak{g}}}\hat{\phi})^2)\\\nonumber
&=&\int  \frac{dk}{2\pi}\frac{\omega_{k}}{{4}}    (a^{\mathfrak{g},\dagger}_ka^{\mathfrak{g}}_{k}+a^{\mathfrak{g}}_ka^{\mathfrak{g},\dagger}_{k}),
\end{eqnarray}
where we used 

\begin{eqnarray}
\int_{x_{\text{h}}}^{x_{\text{h}}+\delta}  dx \frac{ (\hat{\pi}_\phi(x))^2}{h_{\mathfrak{g}}(x)}
&=&-\int  \frac{dk}{2\pi}{\frac{\omega_{k}}{4}}   (a^{\mathfrak{g}}_ka^{\mathfrak{g}}_{-k}-a^{\mathfrak{g},\dagger}_ka^{\mathfrak{g}}_{k}-a^{\mathfrak{g}}_ka^{\mathfrak{g},\dagger}_{k}+a^{\mathfrak{g},\dagger}_{k}a^{\mathfrak{g},\dagger}_{-k}),
\end{eqnarray}
\begin{eqnarray}
\int_{x_{\text{h}}}^{x_{\text{h}}+\delta}  dx h_{\mathfrak{g}}(x) (\partial_{\check{x}^\ast_{\mathfrak{g}}}\hat{\phi}(\check{x}^\ast_{\mathfrak{g}}))^2
&=&-\int  \frac{dk}{2\pi}\frac{\omega_{k}}{{4}}    (-a^{\mathfrak{g}}_ka^{\mathfrak{g}}_{-k}-a^{\mathfrak{g},\dagger}_ka^{\mathfrak{g}}_{k}-a^{\mathfrak{g}}_ka^{\mathfrak{g},\dagger}_{k}-a^{\mathfrak{g},\dagger}_{k}a^{\mathfrak{g},\dagger}_{-k}),
\end{eqnarray}
and the approximation
\begin{eqnarray}
\partial_{\check{x}^\ast_{\mathfrak{g}}}\hat{\phi}(\check{x}^\ast_{\mathfrak{g}})&=&\int \frac{dk}{2\pi}\frac{\mathbf{i}k}{\sqrt{2\omega_k}} (\frac{e^{\mathbf{i}k\check{x}^\ast_{\mathfrak{g}}}}{x}a^{\mathfrak{g}}_k-\frac{e^{-\mathbf{i}k\check{x}^\ast_{\mathfrak{g}}}}{x}a^{\mathfrak{g},\dagger}_k)-\frac{1}{x}\frac{\partial x}{\partial \check{x}^\ast_{\mathfrak{g}}}\hat{\phi}(\check{x}^\ast_{\mathfrak{g}})\\\nonumber
&\approx& \int \frac{dk}{2\pi}\frac{\mathbf{i}k}{\sqrt{2\omega_k}} (\frac{e^{\mathbf{i}k\check{x}^\ast_{\mathfrak{g}}}}{x}a^{\mathfrak{g}}_k-\frac{e^{-\mathbf{i}k\check{x}^\ast_{\mathfrak{g}}}}{x}a^{\mathfrak{g},\dagger}_k).
\end{eqnarray}
Further, by choosing the normal ordering the operators, one has 
\begin{eqnarray}\label{AKAK}
\int_{x_{\text{h}}}^{x_{\text{h}}+\delta}  dx \left(\left.\frac{\partial_xE^x}{2E^\varphi}\right|_{\mathfrak{g}}\right)\hat{H}^\phi(\mathfrak{g})&=&\int  \frac{dk}{2\pi}{\omega_{k}}  a^{\mathfrak{g},\dagger}_ka_{k}.
\end{eqnarray}

 Now, by using Eq.\eqref{AKAK}, the right-hand side of Eq.  \eqref{cons444} can be represented  in the Hilbert space $\mathcal{H}^\phi_{\mathfrak{g}}$ directly, which leads to 
\begin{eqnarray}\label{cons555}
{ M}_{v'}(\mathfrak{g})-{ M}_{v}(\mathfrak{g})=\frac{\int  \frac{dk}{2\pi}{\omega_{k}} \langle\phi'| a^{\mathfrak{g},\dagger}_ka^{\mathfrak{g}}_{k}|\phi\rangle}{\langle\phi'|\phi\rangle}.
\end{eqnarray}
It is easy to see that the solution of  Eq.\eqref{cons555}
 is given by 
 \begin{eqnarray}\label{conssolu}
|\phi\rangle=|\vec{k}_{\vec{n}}(\mathfrak{g})\rangle:=\left(\prod_{{k} }\frac{(a^\dagger_{k})^{n_{k}}}{(\mathcal{N})^{n_k}\sqrt{n_k!}}\right)|0_{\text{B}}\rangle,
\end{eqnarray}
where $\vec{n}=(n_{k_1},n_{k_2},n_{k_3},...)$ with the sequence $k_1<k_2<k_3<...$ composed of all possible values in the spectrum of $k$, $n_{k_1},n_{k_2},n_{k_3},...\in\{0,1,2,3,...\}$ denote the particle number in  modes $k_1,k_2,k_3,...$, $\mathcal{N}:=\frac{1}{\sqrt{2\pi\hbar \delta(k-k')|_{k=k'}}}$ is a normalization constant and $\vec{k}_{\vec{n}}(\mathfrak{g})$ is defined by
\begin{eqnarray}\label{kng}
\vec{k}_{\vec{n}}(\mathfrak{g}):=(\overbrace{({k}_1,...,{k}_1)}^{ n_{k_1}\text{-tuple}\ {k}_1},\overbrace{({k}_2,...,{k}_2)}^{n_{k_2}\text{-tuple}\ {k}_2},\overbrace{({k}_3,...,{k}_3)}^{ n_{k_3}\text{-tuple}\ {k}_3},...)
\end{eqnarray}
  with $(n_{k_1},n_{k_2},n_{k_3},...)$ and $(k_1,k_2,k_3,...)$ satisfying
 \begin{eqnarray}\label{MMomega00}
{ M}_{v'}(\mathfrak{g})-{ M}_{v}(\mathfrak{g})= n_{k_1}\omega_{k_1(\mathfrak{g})}+ n_{k_2}\omega_{k_2(\mathfrak{g})}+ n_{k_3}\omega_{k_3(\mathfrak{g})}+....=:\omega_{\vec{k}_{\vec{n}}(\mathfrak{g})}.
\end{eqnarray}

Here, it is worth to emphasize that, in order to get the Eqs .\eqref{AKAK} and \eqref{cons555}, we use the modified tortoise coordinate $\check{x}^\ast_{\mathfrak{g}}=x^\ast_{\mathfrak{g}}+L-\delta^\ast_{\mathfrak{g}}$ and take the limit $L\to \infty$; These constructions ensure that the Fock states $|\vec{k}_{\vec{n}}\rangle $ defined in this coordinate  $\check{x}^\ast_{\mathfrak{g}}$  are the eigenstate of the operator $\widehat{\tilde{H}^\delta_{\phi}}(\mathfrak{g};v,v')$ on the right-hand side of  Eq. \eqref{cons444}. In fact, this modified tortoise coordinate can only be  defined near a horizon up to now, and thus we only solve a subset of the effective Hamiltonian
constraint   \eqref{cons444}. 

\begin{figure}[htbp]
    \centering 
  \includegraphics[width=0.8\linewidth]{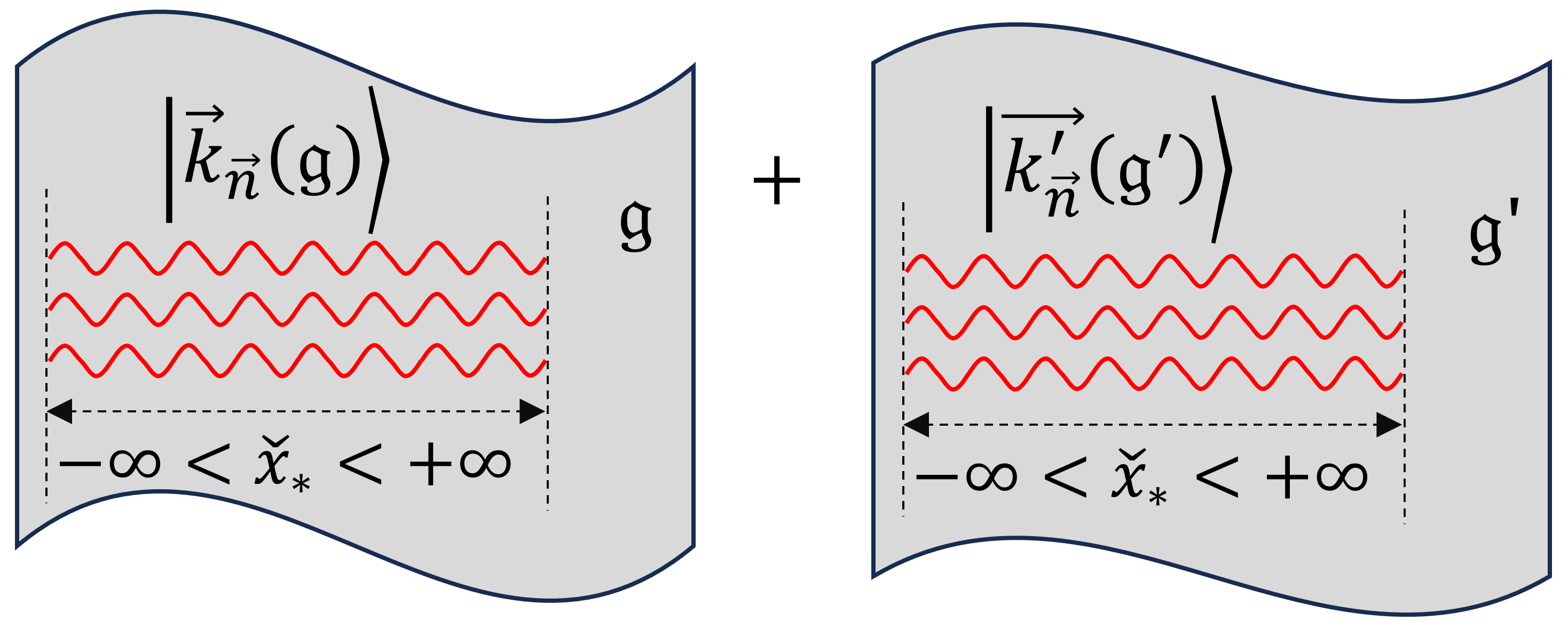}
    \caption{The Fock states $|\vec{k}_{\vec{n}}(\mathfrak{g})\rangle$ and $|\vec{k}'_{\vec{n}}(\mathfrak{g}')\rangle$ (illustrated by red wavy lines) constructed on the classical geometries $\mathfrak{g}$ and $\mathfrak{g}'$ (illustrated by grey surfaces) respectively. The superposition state $|\vec{k}_{\vec{n}}(\mathfrak{g})\rangle+|\vec{k}'_{\vec{n}}(\mathfrak{g}')\rangle$ is ill-defined because of the following two reasons; First,   $|\vec{k}_{\vec{n}}(\mathfrak{g})\rangle$ and $|\vec{k}'_{\vec{n}}(\mathfrak{g}')\rangle$ are constructed on different classical geometries  $\mathfrak{g}$ and $\mathfrak{g}'$  so that they belong to different Hilbert spaces $\mathcal{H}^\phi_{\mathfrak{g}}$ and $\mathcal{H}^\phi_{\mathfrak{g}'}$, while one cannot guarantee a unitary equivalence $\mathcal{H}_{\mathfrak{g}}^\phi\cong \mathcal{H}_{\mathfrak{g}'}^\phi$ for arbitrary $\mathfrak{g}\neq \mathfrak{g}'$; Second,   $|\vec{k}_{\vec{n}}(\mathfrak{g})\rangle$ and $|\vec{k}'_{\vec{n}}(\mathfrak{g}')\rangle$ must be associated to  the classical geometries $\mathfrak{g}$ and $\mathfrak{g}'$ respectively by the Hamiltonian constraint, while the classical geometries can not be superposed.}  \label{superposition classical}
\end{figure}

However, the solution 
\eqref{conssolu} of the effective Hamiltonian constraint equation \eqref{cons444} reveals a fundamental issue inherent in QFT formulated on a fixed classical geometry.    In detail,  for a given spatial geometry $\mathfrak{g}$, the  solutions of Eq. \eqref{cons444}  are certain   Fock states $|\vec{k}_{\vec{n}}(\mathfrak{g})\rangle$ constructed in the modified tortoise coordinate $\check{x}^\ast_{\mathfrak{g}}$ and satisfying Eq.\eqref{MMomega00}. Consider two such Fock states $|\vec{k}_{\vec{n}}(\mathfrak{g})\rangle$ and $|\vec{k}'_{\vec{n}}(\mathfrak{g}')\rangle$, each solving the Hamiltonian constraint \eqref{cons444} and built upon distinct classical geometries $\mathfrak{g}$ and $\mathfrak{g}'$, respectively. Since  $|\vec{k}_{\vec{n}}(\mathfrak{g})\rangle$ and $|\vec{k}'_{\vec{n}}(\mathfrak{g}')\rangle$ are physical states in the sense of weakly solving the Hamiltonian constraint \eqref{cons444}. Hence, it is reasonable to expect that their linear superposition
 \begin{eqnarray}\label{superposition}
|\vec{k}_{\vec{n}}(\mathfrak{g})\rangle+|\vec{k}'_{\vec{n}}(\mathfrak{g}')\rangle \  
\end{eqnarray}
should also represent a valid physical state. However, this superposition is, in fact, \textbf{ill-defined}, and thus  highlights a core limitation of the QFT on a fixed classical background geometry.
 
In detail, the illness of this superposition is caused by
two interrelated reasons:
\begin{itemize}\label{issuereason}
     \item \textbf{The underlying superposition of classical geometry:} Each Fock state $|\vec{k}_{\vec{n}}(\mathfrak{g})\rangle$ is inextricably tied to its background  geometry $\mathfrak{g}$ through the Hamiltonian constraint \eqref{cons444}. Hence, the superposition of states  $|\vec{k}_{\vec{n}}(\mathfrak{g})\rangle$ and $|\vec{k}'_{\vec{n}}(\mathfrak{g}')\rangle$ on distinct background geometries causes the underlying superposition of classical geometry, which leads to this superposition being ill-defined.
    \item \textbf{Inequivalent Fock space}: The states $|\vec{k}_{\vec{n}}(\mathfrak{g})\rangle$ and $|\vec{k}'_{\vec{n}}(\mathfrak{g}')\rangle$  are defined with respect to different classical geometries  $\mathfrak{g}$ and $\mathfrak{g}'$   and therefore reside in different Fock spaces  $\mathcal{H}^\phi_{\mathfrak{g}}$ and $\mathcal{H}^\phi_{\mathfrak{g}'}$, respectively. However, one cannot guarantee an unitary equivalence $\mathcal{H}_{\mathfrak{g}}^\phi\cong \mathcal{H}_{\mathfrak{g}'}^\phi$ for arbitrary $\mathfrak{g}\neq \mathfrak{g}'$,  which leads the superposition of states from these distinct spaces mathematically undefined.
  
\end{itemize}

Our result reveals a key conceptual shortcoming of QFT on a fixed classical spacetime: it cannot consistently accommodate quantum superpositions that involve changes of the background geometry. Resolving this issue typically requires moving beyond fixed backgrounds—toward a fully quantum theory of gravity, where geometry itself becomes a dynamical quantum variable.

\section{Quantum scalar field on semiclassical  geometry  }\label{sec3}

\subsection{ Fock representations on quantum geometry}\label{sec301}

The issue of  QFT on a fixed classical geometry suggests us to consider  QFT on the quantum geometry of spacetime. Specifically, to tackle the illness of  superposition \eqref{superposition}  proposed  in  \ref{issuereason}, one needs to consider the quantum states that satisfy the  following two conditions:
\begin{itemize}
    \item The background geometry sector should be described by some quantum states, i.e. the coherent states  $| \mathfrak{g}\rangle$ of spatial geometry  sharply peaked at the phase space point $\mathfrak{g}$, and now the Fock space $\mathcal{H}_{\mathfrak{g}}^\phi$ can be defined on the semiclassical geometry $| \mathfrak{g}\rangle$.

    \item  The superposition of the product states  $| \mathfrak{g},\phi \rangle:=|\mathfrak{g}\rangle\otimes|\phi\rangle$ should be defined in a certain state space $\tilde{\mathcal{H}}_{\text{g}}\otimes \mathcal{H}^{\phi} $ with $|\mathfrak{g}\rangle\in \tilde{\mathcal{H}}_{\text{g}}$  and $|\phi\rangle\in \mathcal{H}^{\phi} $; Especially, the space $ \mathcal{H}^{\phi}$ must satisfy the unitary equivalence $\mathcal{H}^{\phi}\cong \mathcal{H}_{\mathfrak{g}}^\phi$ of representations  for arbitrary $|\mathfrak{g}\rangle\in \tilde{\mathcal{H}}_{\text{g}}$ , so that the superposition of such states $| \mathfrak{g},\phi \rangle$ is well-defined.
\end{itemize}

Now, let us start by constructing the product state $| \mathfrak{g},\phi \rangle $ and the certain state space $\tilde{\mathcal{H}}_{\text{g}}\otimes \mathcal{H}^{\phi} $ step by step.
We first focus on the gravitational sector. The quantization of the gravitational sector of this system can be realized by following the standard loop quantization method for the spherically symmetric model, which leads to the Hilbert space $\mathcal{H}_{\text{g}}$ of the gravitational sector \cite{Long0propagator,Longrepresentationtwisted}. Then,  the coherent state in  loop quantum theory can be constructed correspondingly (see the details in App.\ref{app1}). The resulting  coherent states $| \mathfrak{g}\rangle\in \mathcal{H}_{\text{g}}$  sharply peaked at the phase space point $\mathfrak{g}$; Specifically,  the coherent states distinguish the classical geometries  $\mathfrak{g}$ and  $\mathfrak{g}'$ by the exponentially decayed inner product  $\langle  \mathfrak{g} | \mathfrak{g}' \rangle$, which satisfies
 \begin{eqnarray}\label{overlap222}
\langle \mathfrak{g}  |\mathfrak{g}' \rangle
\sim \exp(|\mathfrak{g}  -\mathfrak{g}'|^2/t)
\end{eqnarray}
with $|\mathfrak{g}  -\mathfrak{g}'|$ being the distance of the  points  $\mathfrak{g}$ and  $\mathfrak{g}'$ in the phase space,  and  $t\propto G\hbar$ being a controlled parameter labeling the width of the wave packet of the coherent state.  
Moreover, the sharply peaked  coherent states  $| \mathfrak{g}\rangle$  satisfy  the Ehrenfest property, which is given as
\begin{eqnarray}\label{gexpect002}
\langle \mathfrak{g}  |\hat{O}|\mathfrak{g}' \rangle
=  \langle \mathfrak{g}  |\mathfrak{g}' \rangle O(\mathfrak{g}')(1+\mathcal{O}(t)),
\end{eqnarray}
where  $O(\mathfrak{g}')$ is a regular functional on $\mathcal{P}^{\text{g}}$ and   $\hat{O}$ is the operator corresponding to  $O(\mathfrak{g}')$. Consequently, the coherent state $| \mathfrak{g}\rangle$ has a well-behaved semiclassical property and therefore  can be regarded as the semiclassical state.

Let us then turn to the scalar field sector to construct a whole product space $\tilde{\mathcal{H}}_{\text{g}}\otimes \mathcal{H}^{\phi} \subset {\mathcal{H}}_{\text{g}}\otimes \mathcal{H}^{\phi} $. For a given semiclassical  coherent state $| \mathfrak{g}\rangle\in \mathcal{H}_{\text{g}}$ with $\mathfrak{g}\in \mathcal{S}_{ v,v'}$, it is direct to introduce the corresponding Fock space $\mathcal{H}_{\mathfrak{g}}^\phi$ on the geometry $\mathfrak{g}$.  However, the Fock space $\mathcal{H}_{\mathfrak{g}}^\phi$ is associated with certain semiclassical geometry described by  $| \mathfrak{g}\rangle$, and one  cannot ensure a unitary equivalence $\mathcal{H}_{\mathfrak{g}}^\phi\cong \mathcal{H}_{\mathfrak{g}'}^\phi$ of representations  for arbitrary $| \mathfrak{g}\rangle\neq| \mathfrak{g}'\rangle$.  Hence, at this stage  the  Hilbert space $\mathcal{H}^\phi$ cannot be defined  as  $\mathcal{H}^\phi\equiv\mathcal{H}_{\mathfrak{g}}^\phi$ . Nevertheless, we can focus on the subspace $\mathcal{U}_{ v,v'}\subset \mathcal{S}_{ v,v'}$, which  ensures the  unitary equivalence
\begin{eqnarray}\label{isomor}
\mathcal{H}_{\mathfrak{g}}^\phi\cong \mathcal{H}_{\mathfrak{g}'}^\phi,\quad \forall \   \mathfrak{g}, \mathfrak{g}'\in \mathcal{U}_{ v,v'}.
\end{eqnarray} 
Correspondingly, one has the space  $\tilde{\mathcal{H}}_{\text{g}}\otimes \mathcal{H}^{\phi}$ with
\begin{eqnarray}\label{wellspace}
\tilde{\mathcal{H}}_{\text{g}}:=\{| \mathfrak{g}\rangle\in \mathcal{H}_{\text{g}}|\mathfrak{g}\in \mathcal{U}_{ v,v'}\}\ \ \text{and}\ \   \mathcal{H}^\phi\equiv\mathcal{H}_{\mathfrak{g}}^\phi\cong \mathcal{H}_{\mathfrak{g}'}^\phi,\ \ \forall \   \mathfrak{g}, \mathfrak{g}'\in \mathcal{U}_{ v,v'}.
\end{eqnarray} 

The remaining task is to find the subspace $\mathcal{U}_{ v,v'}\subset \mathcal{S}_{ v,v'}$ that satisfies Eq.\eqref{isomor}. Notice that $\mathcal{H}_{\mathfrak{g}}^\phi$ and $ \mathcal{H}_{\mathfrak{g}'}^\phi$  are generated by $(a_k^{\mathfrak{g}},a_k^{\mathfrak{g},\dagger})$ and $(a_k^{\mathfrak{g}'},a_k^{\mathfrak{g}',\dagger})$ respectively, and thus one has the Bogoliubov transformation mapping the annihilation operators $a_k^{\mathfrak{g}'}$  to 
 \begin{eqnarray}\label{}
a_k^{\mathfrak{g}}=\int \frac{dj}{2\pi}(\alpha_{kj}(\mathfrak{g}',\mathfrak{g})a_j^{\mathfrak{g}'}+\beta_{kj}(\mathfrak{g}',\mathfrak{g})a_j^{\mathfrak{g}',\dagger}).
\end{eqnarray}
 Recalling Eqs.\eqref{ak}, \eqref{piphiaa}, and \eqref{phiaa}, one can immediately get  \begin{eqnarray}\label{}
\alpha_{kk'}(\mathfrak{g}',\mathfrak{g})=\frac{1}{2}\sqrt{\frac{\omega_{k}}{\omega_{k'}}}(\lim_{L\to\infty}\int_{-L}^{+L} d\check{x}^\ast_{\mathfrak{g}} {e^{-\mathbf{i}k\check{x}^\ast_{\mathfrak{g}}}}{e^{\mathbf{i}k'\check{x}^\ast_{\mathfrak{g}'}}})+\frac{1}{2}\sqrt{\frac{\omega_{k'}}{\omega_{k}}}(\lim_{L\to\infty}\int_{-L}^{+L} d\check{x}^\ast_{\mathfrak{g}'} {e^{-\mathbf{i}k\check{x}^\ast_{\mathfrak{g}}}}{e^{\mathbf{i}k'\check{x}^\ast_{\mathfrak{g}'}}})
\end{eqnarray}
and 
 \begin{eqnarray}\label{}
\beta_{kk'}(\mathfrak{g}',\mathfrak{g})=\frac{1}{2}\sqrt{\frac{\omega_{k}}{\omega_{k'}}}(\lim_{L\to\infty}\int_{-L}^{+L} d\check{x}^\ast_{\mathfrak{g}} {e^{-\mathbf{i}k\check{x}^\ast_{\mathfrak{g}}}}{e^{-\mathbf{i}k'\check{x}^\ast_{\mathfrak{g}'}}})-\frac{1}{2}\sqrt{\frac{\omega_{k'}}{\omega_{k}}}(\lim_{L\to\infty}\int_{-L}^{+L} d\check{x}^\ast_{\mathfrak{g}'} {e^{-\mathbf{i}k\check{x}^\ast_{\mathfrak{g}}}}{e^{-\mathbf{i}k'\check{x}^\ast_{\mathfrak{g}'}}}).
\end{eqnarray}
Then, by using the Shale–Stinespring criterion, one has the following result;
\begin{itemize}
    \item The unitary equivalence  $\mathcal{H}_{\mathfrak{g}}^\phi\cong \mathcal{H}_{\mathfrak{g}'}^\phi$  holds if and only if the Bogoliubov transformation mapping the annihilation operators $a_k^{\mathfrak{g}'}$  to 
 \begin{eqnarray}\label{Bogo}
a_k^{\mathfrak{g}}=\int \frac{dj}{2\pi}(\alpha_{kj}(\mathfrak{g}',\mathfrak{g})a_j^{\mathfrak{g}'}+\beta_{kj}(\mathfrak{g}',\mathfrak{g})a_j^{\mathfrak{g}',\dagger})
\end{eqnarray}
 can be implemented by a unitary operator $U_{\mathfrak{g},\mathfrak{g}'}$ on the Fock space, i.e., there exists a unitary  $U_{\mathfrak{g}',\mathfrak{g}}$  such that $a_k^{\mathfrak{g}}=U_{\mathfrak{g}',\mathfrak{g}}a_k^{\mathfrak{g}'}U_{\mathfrak{g}',\mathfrak{g}}^\dagger$ for all $k$; Moreover, such a unitary  $U_{\mathfrak{g}',\mathfrak{g}}$ exists  if and only if the matrix of Bogoliubov coefficients $\beta_{kj}(\mathfrak{g}',\mathfrak{g})$ is a Hilbert–Schmidt operator, namely,
 \begin{eqnarray}\label{betalessinfty}
\int dkdj| \beta_{kj}(\mathfrak{g}',\mathfrak{g})|^2< \infty.
\end{eqnarray}

\end{itemize}
Now, the subspace $\mathcal{U}_{ v,v'}\subset \mathcal{S}_{ v,v'}$ that satisfies Eq.\eqref{isomor} can be defined by
\begin{eqnarray}
\mathcal{U}_{ v,v'}:=\{\mathfrak{g}''\in\mathcal{S}_{ v,v'}|\int dkdj| \beta_{kj}(\mathfrak{g}',\mathfrak{g})|^2< \infty, \  \forall \mathfrak{g}',\mathfrak{g}\in \mathcal{U}_{ v,v'} \}.
\end{eqnarray} 
Moreover, further discussion of the subspace $\mathcal{U}_{ v,v'}\subset \mathcal{S}_{ v,v'}$ based on an example is given in App.\ref{app2}.

\subsection{The weak solution of quantum Hamiltonian constraint}
\label{sec302}
Now, the illness of  superposition \eqref{superposition}  proposed  in  \ref{issuereason} disappears in the quantum state space $\tilde{\mathcal{H}}_{\text{g}}\otimes \mathcal{H}^{\phi}$ defined by \eqref{wellspace}.
Hence, it is ready to consider
the totally quantized version of  Eq.\eqref{cons111}, which reads
\begin{equation}\label{cons222}
\langle \hat{M}_{v'}\rangle-\langle \hat{M}_{v}\rangle=\langle\widehat{\tilde{H}^\delta_{\phi}}(v,v')\rangle,
\end{equation}
where $\hat{M}_{v}, \hat{M}_{v'}$ is the operator of ${M}_{v}, {M}_{v'}$, $\widehat{\tilde{H}^\delta_{\phi}}(v,v')$ is the operator of ${\tilde{H}^\delta_{\phi}}(v,v')$, and $\langle\hat{P}\rangle$ represents the expectation value of an operator $\hat{P}$.
For an arbitrary state $| \mathfrak{g},\phi \rangle\in  \tilde{\mathcal{H}}_{\text{g}}\otimes \mathcal{H}^{\phi}$,  Eq.\eqref{cons222}  can be written as 
\begin{equation}
\frac{\langle \mathfrak{g}',\phi'|\hat{M}_{v'}|\mathfrak{g},\phi\rangle}{\langle \mathfrak{g}',\phi'|\mathfrak{g},\phi\rangle}-\frac{\langle  \mathfrak{g}',\phi'|\hat{M}_{v}|\mathfrak{g},\phi\rangle}{{\langle \mathfrak{g}',\phi'|\mathfrak{g},\phi\rangle}}=\frac{\langle \mathfrak{g}',\phi'|\widehat{\tilde{H}^\delta_{\phi}}(v,v')|\mathfrak{g},\phi \rangle}{{\langle \mathfrak{g}',\phi'|\mathfrak{g},\phi\rangle}}.
\end{equation}
Furthermore, by using the property \eqref{gexpect002}
 of the coherent state  $|\mathfrak{g}\rangle$, one has
\begin{eqnarray}\label{gexpect111}
\langle \mathfrak{g}' ,\phi' |(\hat{ M}_{v'}-\hat{ M}_{v})|\mathfrak{g} ,\phi \rangle
=\langle \mathfrak{g}',\phi'  |\mathfrak{g},\phi  \rangle({ M}_{v'}(\mathfrak{g})-{ M}_{v}(\mathfrak{g}))(1+\mathcal{O}(t))
\end{eqnarray}
and 
\begin{eqnarray}\label{gexpect222}
\langle \mathfrak{g}'  ,\phi'|\widehat{\tilde{H}^\delta_{\phi}}(v,v')| \mathfrak{g},\phi \rangle=\langle\phi'|\widehat{\tilde{H}^\delta_{\phi}}(\mathfrak{g};v,v')|\phi\rangle\langle \mathfrak{g}'  |\mathfrak{g} \rangle(1+\mathcal{O}'(t)),
\end{eqnarray}
where ${ M}_{v'}(\mathfrak{g}), { M}_{v}(\mathfrak{g})$ and $\widehat{\tilde{H}^\delta_{\phi}}(\mathfrak{g};v,v')$ are the values of ${ M}_{v'}, { M}_{v}$ and $\widehat{\tilde{H}^\delta_{\phi}}(v,v')$ at the phase space point labeled by $\mathfrak{g}$, with $\widehat{\tilde{H}^\delta_{\phi}}(\mathfrak{g};v,v')$ being given by Eq.\eqref{Hgvv}.

Now, Eq.\eqref{cons222}  can be simplified as
\begin{eqnarray}\label{MMgg222}
{ M}_{v'}(\mathfrak{g})-{ M}_{v}(\mathfrak{g})=\frac{\langle\phi'|\widehat{\tilde{H}^\delta_{\phi}}(\mathfrak{g};v,v')|\phi\rangle}{\langle\phi'|\phi\rangle}
\end{eqnarray}
at the leading order of $t$. It is easy to see that  Eq.\eqref{MMgg222}
 is identical to Eq.\eqref{cons444}. Hence, by recalling the solution \eqref{conssolu} of  Eq.\eqref{cons444},  a weak solution of the quantum constraint  \eqref{cons222} can be given by 
\begin{eqnarray}\label{conssolu222}
|\mathfrak{g},\phi\rangle=|\mathfrak{g},\vec{k}_{\vec{n}}(\mathfrak{g})\rangle\equiv|\mathfrak{g}\rangle\otimes|\vec{k}_{\vec{n}}(\mathfrak{g})\rangle  
\end{eqnarray}
where
$\vec{k}_{\vec{n}}(\mathfrak{g}) $ is given by Eq.\eqref{kng} for the specific $\mathfrak{g}$. 
Equivalently, one can also express the solution  \eqref{conssolu222} as 
 \begin{eqnarray}\label{conssolu333}
|\mathfrak{g},\phi\rangle=|\mathfrak{g}(\vec{k}_{\vec{n}}),\vec{k}_{\vec{n}}\rangle\equiv|\mathfrak{g}(\vec{k}_{\vec{n}})\rangle\otimes|\vec{k}_{\vec{n}}\rangle  
\end{eqnarray}
where 
\begin{eqnarray}\label{kn222}
\vec{k}_{\vec{n}}:=(\overbrace{({k}_1,...,{k}_1)}^{ n_{k_1}\text{-tuple}\ {k}_1},\overbrace{({k}_2,...,{k}_2}^{n_{k_2}\text{-tuple}\ {k}_2},\overbrace{({k}_3,...,{k}_3)}^{ n_{k_3}\text{-tuple}\ {k}_3},...)
\end{eqnarray}
with $\mathfrak{g}(\vec{k}_{\vec{n}})\in \mathcal{U}_{ v,v'}$ being the geometry satisfying
 \begin{eqnarray}
{ M}_{v'}(\mathfrak{g}(\vec{k}_{\vec{n}}))-{ M}_{v}(\mathfrak{g}(\vec{k}_{\vec{n}}))= n_{k_1}\omega_{k_1}+ n_{k_2}\omega_{k_2}+n_{k_3} \omega_{k_3}+...=:\omega_{\vec{k}_{\vec{n}}}
\end{eqnarray}
for specific  $(n_{k_1},n_{k_2},n_{k_3},...)$ and $(k_1,k_2,k_3,...)$.

More generally, the weak solutions of the quantum Hamiltonian constraint \eqref{cons222}  in  space $\tilde{\mathcal{H}}_{\text{g}}\otimes \mathcal{H}^{\phi}$ are given by the linear combinations of such states \eqref{conssolu222} or \eqref{conssolu333}, e.g. the well-defined superposition state 
\begin{eqnarray}\label{superposition333}
c|\mathfrak{g}(\vec{k}_{\vec{n}})\rangle\otimes|\vec{k}_{\vec{n}}\rangle +c'|\mathfrak{g}(\vec{k}'_{\vec{n}})\rangle\otimes|\vec{k}'_{\vec{n}}\rangle+c''|\mathfrak{g}(\vec{k}''_{\vec{n}})\rangle\otimes|\vec{k}''_{\vec{n}}\rangle+... \in  \tilde{\mathcal{H}}_{\text{g}}\otimes \mathcal{H}^{\phi}.
\end{eqnarray}
In other words, the weak solution of the quantum Hamiltonian constraint \eqref{cons222}  in the space $\tilde{\mathcal{H}}_{\text{g}}\otimes \mathcal{H}^{\phi}$ spans a subspace $(\tilde{\mathcal{H}}_{\text{g}}\otimes \mathcal{H}^{\phi})^{\text{sol.}}_{v,v'}$ defined by
\begin{eqnarray}
(\tilde{\mathcal{H}}_{\text{g}}\otimes \mathcal{H}^{\phi})^{\text{sol.}}_{v,v'}:=\{|\mathfrak{g}(\vec{k}_{\vec{n}})\rangle\otimes|\vec{k}_{\vec{n}}\rangle \in  \tilde{\mathcal{H}}_{\text{g}}\otimes \mathcal{H}^{\phi}|{ M}_{v'}(\mathfrak{g}(\vec{k}_{\vec{n}}))-{ M}_{v}(\mathfrak{g}(\vec{k}_{\vec{n}}))=\omega_{\vec{k}_{\vec{n}}}\}.
\end{eqnarray}

It is worth to have some discussions on the states in $(\tilde{\mathcal{H}}_{\text{g}}\otimes \mathcal{H}^{\phi})^{\text{sol.}}_{v,v'}$.
\begin{enumerate}
    \item Notice that we only considered the case that  $v$ and $ v'$ satisfy $x(v)=x_{\text{h}}$ and $x(v')=x_{\text{h}}+\delta$ in Eq.  \eqref{cons444}. Hence,
the states in $(\tilde{\mathcal{H}}_{\text{g}}\otimes \mathcal{H}^{\phi})^{\text{sol.}}_{v,v'}$ are just some weak solutions of a subset of the quantum Hamiltonian constraints  \eqref{cons222}  at the order of  $t$.
    \item Although the states in $(\tilde{\mathcal{H}}_{\text{g}}\otimes \mathcal{H}^{\phi})^{\text{sol.}}_{v,v'}$ are just some weak solutions of a subset of the quantum Hamiltonian constraints  \eqref{cons222} at the order of  $t$, it does indicate the entanglement between the quantum matter field and geometry. Especially,   the weak solutions in $(\tilde{\mathcal{H}}_{\text{g}}\otimes \mathcal{H}^{\phi})^{\text{sol.}}_{v,v'}$ can be regarded as the physical states of quantum gravity in the specific finite region $x(v)<x\leq x(v')$ approximately; Hence, once we focus on this  region, it is reasonable to require that all of the states discussed in the QFTCS  should effectively be  approximate quantum states of the linear combination of  physical states in $(\tilde{\mathcal{H}}_{\text{g}}\otimes \mathcal{H}^{\phi})^{\text{sol.}}_{v,v'}$. In the next section, we  discuss the Hartle-Hawking vacuum as an example.
\end{enumerate}

\subsection{Generalized Hartle-Hawking vacuum on quantum geometry}\label{sec303}
As mentioned above, the issue of  QFTCS can be resolved by considering the  state in $(\tilde{\mathcal{H}}_{\text{g}}\otimes \mathcal{H}^{\phi})^{\text{sol.}}_{v,v'}$ .
Specifically, once we consider the states in  QFTCS in the region $x(v)<x<x(v')$, we should make the substitution 
\begin{eqnarray}\label{substi}
|\vec{k}_{\vec{n}}\rangle _{\mathfrak{g}} \rightarrow |\mathfrak{g}(\vec{k}_{\vec{n}}),\vec{k}_{\vec{n}}\rangle\in (\tilde{\mathcal{H}}_{\text{g}}\otimes \mathcal{H}^{\phi})^{\text{sol.}}_{v,v'},
\end{eqnarray}
where $|\vec{k}_{\vec{n}}\rangle _{\mathfrak{g}} $ defined in App.\ref{app3} is the Fock state constructed on some fixed classical geometry $\mathfrak{g}\in\mathcal{S}_{ v,v'}$.
As an example, one can construct the generalized Hartle-Hawking vacuum on quantum geometry by using   substitution  \eqref{substi}.  In detail,  on the fixed classical geometry $\mathfrak{g}\in\mathcal{S}_{ v,v'}$, the local Hartle-Hawking vacuum in the region $x(v)<x<x(v')$ in QFTCS   is given as (see  details in App.\ref{app3})
\begin{equation}\label{0Kk}
|0_{\text{HH}} \rangle_{\mathfrak{g}}= \sum_{\vec{n} }\left(\prod_{{k} }{(1-e^{-2\pi\xi_{\mathfrak{g}}^{-1}\omega_{{k} }})^{1/2}}e^{-\pi n_{k} \omega_{k}/\xi_{\mathfrak{g}}}\right)|\vec{k}_{\vec{n}}\rangle_{\mathfrak{g}},
\end{equation}
By using  substitution  \eqref{substi}, one obtains the generalized Hartle-Hawking vacuum  on the quantum geometry, which reads

\begin{equation}\label{HHVg}
|0 _{\text{HHQG}}\rangle:=\sum_{\vec{n} }\left(\prod_{{k} \in\mathbb{R}}{\Big(1-\exp({-2\pi\omega_{{k} }/\xi_{\mathfrak{g}(\vec{k}_{\vec{n}})}})\Big)^{1/2}}\exp\big({-\pi n_{k} \omega_{k}/\xi_{\mathfrak{g}(\vec{k}_{\vec{n}})}}\big)\right)|\mathfrak{g}(\vec{k}_{\vec{n}}),\vec{k}_{\vec{n}}\rangle;
\end{equation} 
Specifically,  $|0 _{\text{HHQG}}\rangle$ is the generalized Hartle-Hawking vacuum on quantum geometry in the specific finite region $x(v)<x<x(v')$, where the apparent horizon for each  $\mathfrak{g}(\vec{k}_{\vec{n}})$ is located at $x=x(v)$,  $|\mathfrak{g}(\vec{k}_{\vec{n}})\rangle\otimes|\vec{k}_{\vec{n}}\rangle \in (\tilde{\mathcal{H}}_{\text{g}}\otimes \mathcal{H}^{\phi})^{\text{sol.}}_{v,v'}$   is  the physical state of quantum gravity and $\xi_{\mathfrak{g}(\vec{k}_{\vec{n}})}$ is an undetermined parameter. 

It is necessary to have some discussions on the state $|0 _{\text{HHQG}}\rangle$. First, by comparing  $|0 _{\text{HHQG}}\rangle$ with  $|0_{\text{HH}} \rangle$, one  can see that 
the Fock states $|\vec{k}_{\vec{n}}\rangle$ in the combination    \eqref{HHVg} are entangled with the quantum geometry state $|\mathfrak{g}(\vec{k}_{\vec{n}})\rangle$, while there is no such entanglement in the combination \eqref{0Kk} . Especially,  this entanglement in the state  $|0 _{\text{HHQG}}\rangle$ comes from  weakly solved Hamiltonian constraints  \eqref{cons222}. 
Second, the state $|0 _{\text{HHQG}}\rangle$ is a quantum state for both  spacetime geometry and the scalar field. Thus, it is different from any vacuum states in the QFT on a fixed classical spacetime. In fact, the state $|0 _{\text{HHQG}}\rangle$ is not defined  as an eigenstate of any annihilation operator. In order to determine the parameter   $\xi_{\mathfrak{g}(\vec{k}_{\vec{n}})}$ in Eq.  \eqref{HHVg} , let us require that the state  $|0 _{\text{HHQG}}\rangle$  should get back to the local Hartle-Hawking vacuum  $|0_{\text{HH}} \rangle_{\mathfrak{g}_{\text{c}}}$  in the  semiclassical limit of the quantum geometry; In other words, the parameter   $\xi_{\mathfrak{g}(\vec{k}_{\vec{n}})}$ should be determined by the semiclassical consistency conditions
\begin{equation}\label{paradeter}
\lim_{\epsilon(\mathfrak{g}_{\text{c}},\mathfrak{g}(\vec{k}_{\vec{n}}))\to0}{A^{\mathfrak{g}_{\text{c}}}_k}|0_{\text{HHQG}} \rangle=0,
\end{equation}  
where $\epsilon(\mathfrak{g}_{\text{c}},\mathfrak{g}(\vec{k}_{\vec{n}}))\equiv\sqrt{1-|\langle\mathfrak{g}_{\text{c}}|\mathfrak{g}(\vec{k}_{\vec{n}})\rangle|^2}$, ${A^{\mathfrak{g}_{\text{c}}}_k}, A^{\mathfrak{g}_{\text{c}},\dagger}_k$ are defined by Eq.\eqref{AAdag}  on the classical geometry $\mathfrak{g}_{\text{c}}$, with  $\mathfrak{g}_{\text{c}}$  given by the expectation values
\begin{equation}\label{gexpect}
\mathfrak{g}_{\text{c}}=\{ 
\frac{\langle 0_{\text{HHQG}}|\hat{E}^\varphi|0_{\text{HHQG}} \rangle }{\langle 0_{\text{HHQG}}|0_{\text{HHQG}} \rangle },\frac{\langle 0_{\text{HHQG}}|\hat{K}_\varphi|0_{\text{HHQG}} \rangle }{\langle 0_{\text{HHQG}}|0 _{\text{HHQG}}\rangle }\}.
\end{equation}  
In fact, one can notice that the coefficient $\exp(-\pi n_{k} \omega_{k}/\xi_{\mathfrak{g}(\vec{k}_{\vec{n}})})$  in the superposition \eqref{HHVg}
is exponentially suppressed with $n_k$ going large. Thus,  the calculations of the expectation values in \eqref{gexpect} are dominated by the terms in  \eqref{HHVg} with small $n_k$. 
 For the case that $\mathfrak{g}_{\text{c}}$ corresponds to the geometry of a large mass black hole, the difference  between $\mathfrak{g}_{\text{c}}$ and each $\mathfrak{g}(\vec{k}_{\vec{n}})$ with small $n_k$ in Eq.\eqref{HHVg} is just some small perturbations. 
Hence, one has the estimation   
$\xi_{\mathfrak{g}(\vec{k}_{\vec{n}})}\approx\xi_{\mathfrak{g}_{\text{c}}}=\left.\frac{1}{2}\partial_x(\frac{x^2}{h_{\mathfrak{g}_{\text{c}}}(x)})\right|_{x=x_{\text{h},\mathfrak{g}_{\text{c}}}}$ with $x=x_{\text{h},\mathfrak{g}_{\text{c}}}$ being the apparent horizon in the geometry $\mathfrak{g}_{\text{c}}$.

      It is worth to emphasize that one can consider another type of generalized Hartle-Hawking vacuum state $|0_{\text{HH}} ,\mathfrak{g}_{\text{c}}\rangle:=|0_{\text{HH}} \rangle_{\mathfrak{g}_{\text{c}}}\otimes |\mathfrak{g}_{\text{c}}\rangle$, which   also satisfies some semiclassical consistency conditions 
      \begin{equation}
{A^{\mathfrak{g}_{\text{c}}}_k}|0_{\text{HH}} ,\mathfrak{g}_{\text{c}}\rangle=0,
\end{equation}  
and 
  \begin{equation}
\mathfrak{g}_{\text{c}}=\{ 
\frac{\langle0_{\text{HH}} ,\mathfrak{g}_{\text{c}}|\hat{E}^\varphi|0_{\text{HH}} ,\mathfrak{g}_{\text{c}}\rangle}{\langle0_{\text{HH}} ,\mathfrak{g}_{\text{c}}| 0_{\text{HH}} ,\mathfrak{g}_{\text{c}}\rangle},\frac{\langle0_{\text{HH}} ,\mathfrak{g}_{\text{c}}|\hat{K}_\varphi|0_{\text{HH}} ,\mathfrak{g}_{\text{c}}\rangle}{\langle0_{\text{HH}} ,\mathfrak{g}_{\text{c}}| 0_{\text{HH}} ,\mathfrak{g}_{\text{c}}\rangle }\}.
\end{equation}
Note that both  states $|0_{\text{HH}} ,\mathfrak{g}_{\text{c}}\rangle$ and $|0 _{\text{HHQG}}\rangle$ give the local Hartle-Hawking vacuum  $|0_{\text{HH}} \rangle_{\mathfrak{g}_{\text{c}}}$  in the  semiclassical limit of the quantum geometry by some semiclassical consistency conditions.
Nevertheless, they possess fundamentally different physical interpretations at the quantum geometric level;
 In fact, the state  $|0 _{\text{HHQG}}\rangle$  weakly solves the quantum Hamiltonian constraints  \eqref{cons222}, and these constraints naturally encode the entanglements between field excitations and the underlying quantum geometry; However,  it is obvious that the state $|0_{\text{HH}} ,\mathfrak{g}_{\text{c}}\rangle$ fails to satisfy the quantum Hamiltonian constraints  \eqref{cons222} and contains no geometry-matter entanglement.
This distinction highlights a core advantage of the quantum gravity framework: physical constraints not only select permissible states but also determine their intrinsic entanglement structure. The construction of   $|0 _{\text{HHQG}}\rangle$  embodies the paradigm shift from "QFT {on} curved spacetime" to "QFT {with} quantum spacetime," where geometry is no longer a fixed background but a physical degree of freedom that participates equally with matter fields in quantum dynamics.

\section{Conclusions and Discussion}\label{sec4}

In this paper, we present a consistent framework for describing quantum scalar fields on semiclassical quantum geometries, resolving a foundational inconsistency in QFT on curved spacetime. 
We model the background geometry with semiclassical coherent states from loop quantum gravity.
In particular, we consider the coherent states whose labels are restricted to a subspace $\mathcal{U}_{ v,v'}$ of the phase space of geometry, on which the Fock representations of the scalar field are unitarily equivalent.
This restriction enables the construction of the space $\tilde{\mathcal{H}}_{\text{g}}\otimes \mathcal{H}^{\phi}$, in which superpositions of the geometry-matter product states become well-defined. 
Moreover, we obtain weak solutions to the quantum Hamiltonian constraint of GR in space $\tilde{\mathcal{H}}_{\text{g}}\otimes \mathcal{H}^{\phi}$.  Consequently, general physical states --defined as the linear combination of these weak solutions  in $\tilde{\mathcal{H}}_{\text{g}}\otimes \mathcal{H}^{\phi}$--- are inherently entangled.
We also construct the generalized Hartle–Hawking vacuum on quantum geometry by superposing the weak solutions  in $\tilde{\mathcal{H}}_{\text{g}}\otimes \mathcal{H}^{\phi}$, which produces the state $|0 _{\text{HHQG}}\rangle$. Unlike its QFTCS counterpart, this vacuum is not defined with respect to a fixed background.
More importantly, the state $|0 _{\text{HHQG}}\rangle$ exhibits the inherent entanglement between geometry and matter, arising from the fundamental principles of quantum gravity.

Several important directions remain open for future investigation.
\begin{enumerate}
    \item  The current construction of the subspace $\mathcal{U}_{ v,v'}$  relies on spherical symmetry and a specific region near the apparent horizon. It is essential to investigate whether analogous subspaces exist in more general settings. 
    \item  Our construction provides kinematical states that satisfy the quantum Hamiltonian constraint weakly. A complete description of black hole evaporation would require a notion of time evolution, potentially via relational observables or a path-integral formulation over histories of entangled geometry–matter configurations.
    \item  It would be valuable to compute corrections to the Hawking spectrum arising from the quantum-geometric structure of $|0 _{\text{HHQG}}\rangle$ by considering the case where  $\mathfrak{g}_{\text{c}}$ corresponds to the geometry of a  black hole with small mass, which could provide observational signatures of Planck-scale physics.
    \item The built-in geometry-matter entanglement provides a natural framework for studying the black hole information paradox. Future work should explore whether unitarity is preserved under full quantum gravitational evolution, with entanglement playing a key role in information recovery.
\end{enumerate}

In summary, our work demonstrates that, by treating geometry as quantum variables, we resolve the foundational inconsistencies of QFTCS on a fixed-background and uncover a richer structure in which spacetime and matter co-evolve as entangled quantum entities. This result not only resolves a long-standing conceptual issue, but also  provides a new framework  for understanding the black hole information paradox and the nature of quantum spacetime.

\section*{Acknowledgments}
This work is supported by the project funded by  the National Natural Science Foundation of China (NSFC) (with Nos. 12405062, 12275022, and 12505055), and ``the Fundamental Research Funds for the Central Universities'' (with Nos. 21624340 and 2253100010).

\bibliographystyle{unsrt}

\bibliography{ref}

@article{Hawking:1976ra,
    author = "Hawking, S. W.",
    title = "{Breakdown of Predictability in Gravitational Collapse}",
    doi = "10.1103/PhysRevD.14.2460",
    journal = "Phys. Rev. D",
    volume = "14",
    pages = "2460--2473",
    year = "1976"
}

@article{Bekenstein:1974ax,
    author = "Bekenstein, Jacob D.",
    title = "{Generalized second law of thermodynamics in black hole physics}",
    doi = "10.1103/PhysRevD.9.3292",
    journal = "Phys. Rev. D",
    volume = "9",
    pages = "3292--3300",
    year = "1974"
}

@article{Harlow:2014yka,
    author = "Harlow, Daniel",
    title = "{Jerusalem Lectures on Black Holes and Quantum Information}",
    eprint = "1409.1231",
    archivePrefix = "arXiv",
    primaryClass = "hep-th",
    doi = "10.1103/RevModPhys.88.015002",
    journal = "Rev. Mod. Phys.",
    volume = "88",
    pages = "015002",
    year = "2016"
}

@article{Bekenstein:1973ur,
    author = "Bekenstein, Jacob D.",
    title = "{Black holes and entropy}",
    doi = "10.1103/PhysRevD.7.2333",
    journal = "Phys. Rev. D",
    volume = "7",
    pages = "2333--2346",
    year = "1973"
}

@article{Carlip:2008wv,
    author = "Carlip, Steven",
    editor = "Papantonopoulos, Eleftherios",
    title = "{Black Hole Thermodynamics and Statistical Mechanics}",
    eprint = "0807.4520",
    archivePrefix = "arXiv",
    primaryClass = "gr-qc",
    doi = "10.1007/978-3-540-88460-6_3",
    journal = "Lect. Notes Phys.",
    volume = "769",
    pages = "89--123",
    year = "2009"
}

@article{Wald:1999vt,
    author = "Wald, Robert M.",
    title = "{The thermodynamics of black holes}",
    eprint = "gr-qc/9912119",
    archivePrefix = "arXiv",
    doi = "10.12942/lrr-2001-6",
    journal = "Living Rev. Rel.",
    volume = "4",
    pages = "6",
    year = "2001"
}

@article{Hollands:2014eia,
    author = "Hollands, Stefan and Wald, Robert M.",
    title = "{Quantum fields in curved spacetime}",
    eprint = "1401.2026",
    archivePrefix = "arXiv",
    primaryClass = "gr-qc",
    doi = "10.1016/j.physrep.2015.02.001",
    journal = "Phys. Rept.",
    volume = "574",
    pages = "1--35",
    year = "2015"
}

@book{Birrell_Davies_1982,
place={Cambridge}, series={Cambridge Monographs on Mathematical Physics}, title={Quantum Fields in Curved Space}, publisher={Cambridge University Press}, author={Birrell, N. D. and Davies, P. C. W.}, year={1982}, collection={Cambridge Monographs on Mathematical Physics}}

@book{Wald:1995yp,
    author = "Wald, Robert M.",
    title = "{Quantum Field Theory in Curved Space-Time and Black Hole Thermodynamics}",
    isbn = "978-0-226-87027-4",
    publisher = "University of Chicago Press",
    address = "Chicago, IL",
    series = "Chicago Lectures in Physics",
    year = "1995"
}

@article{Shale1962,
  author  = {Shale, David},
  title   = {Linear Symmetries of Free Boson Fields},
  journal = {Transactions of the American Mathematical Society},
  year    = {1962},
  volume  = {103},
  pages   = {149--167},
  doi     = {10.2307/1993745}
}

@article{Almheiri:2019hni,
    author = "Almheiri, Ahmed and Mahajan, Raghu and Maldacena, Juan and Zhao, Ying",
    title = "{The Page curve of Hawking radiation from semiclassical geometry}",
    eprint = "1908.10996",
    archivePrefix = "arXiv",
    primaryClass = "hep-th",
    doi = "10.1007/JHEP03(2020)149",
    journal = "JHEP",
    volume = "03",
    pages = "149",
    year = "2020"
}

@article{Mathur:2009hf,
    author = "Mathur, Samir D.",
    editor = "Uranga, A. M.",
    title = "{The Information paradox: A Pedagogical introduction}",
    eprint = "0909.1038",
    archivePrefix = "arXiv",
    primaryClass = "hep-th",
    doi = "10.1088/0264-9381/26/22/224001",
    journal = "Class. Quant. Grav.",
    volume = "26",
    pages = "224001",
    year = "2009"
}

@article{Penington:2019npb,
    author = "Penington, Geoffrey",
    title = "{Entanglement Wedge Reconstruction and the Information Paradox}",
    eprint = "1905.08255",
    archivePrefix = "arXiv",
    primaryClass = "hep-th",
    doi = "10.1007/JHEP09(2020)002",
    journal = "JHEP",
    volume = "09",
    pages = "002",
    year = "2020"
}

@article{Unruh:2017uaw,
    author = "Unruh, William G. and Wald, Robert M.",
    title = "{Information Loss}",
    eprint = "1703.02140",
    archivePrefix = "arXiv",
    primaryClass = "hep-th",
    doi = "10.1088/1361-6633/aa778e",
    journal = "Rept. Prog. Phys.",
    volume = "80",
    number = "9",
    pages = "092002",
    year = "2017"
}

@article{Page:2013dx,
    author = "Page, Don N.",
    title = "{Time Dependence of Hawking Radiation Entropy}",
    eprint = "1301.4995",
    archivePrefix = "arXiv",
    primaryClass = "hep-th",
    doi = "10.1088/1475-7516/2013/09/028",
    journal = "JCAP",
    volume = "09",
    pages = "028",
    year = "2013"
}

@article{Page:1993wv,
    author = "Page, Don N.",
    title = "{Information in black hole radiation}",
    eprint = "hep-th/9306083",
    archivePrefix = "arXiv",
    reportNumber = "ALBERTA-THY-24-93",
    doi = "10.1103/PhysRevLett.71.3743",
    journal = "Phys. Rev. Lett.",
    volume = "71",
    pages = "3743--3746",
    year = "1993"
}

@article{Almheiri:2019qdq,
    author = "Almheiri, Ahmed and Hartman, Thomas and Maldacena, Juan and Shaghoulian, Edgar and Tajdini, Amirhossein",
    title = "{Replica Wormholes and the Entropy of Hawking Radiation}",
    eprint = "1911.12333",
    archivePrefix = "arXiv",
    primaryClass = "hep-th",
    doi = "10.1007/JHEP05(2020)013",
    journal = "JHEP",
    volume = "05",
    pages = "013",
    year = "2020"
}

@article{Almheiri:2020cfm,
    author = "Almheiri, Ahmed and Hartman, Thomas and Maldacena, Juan and Shaghoulian, Edgar and Tajdini, Amirhossein",
    title = "{The entropy of Hawking radiation}",
    eprint = "2006.06872",
    archivePrefix = "arXiv",
    primaryClass = "hep-th",
    doi = "10.1103/RevModPhys.93.035002",
    journal = "Rev. Mod. Phys.",
    volume = "93",
    number = "3",
    pages = "035002",
    year = "2021"
}

@article{Hawking:1974rv,
    author = "Hawking, S. W.",
    title = "{Black hole explosions}",
    doi = "10.1038/248030a0",
    journal = "Nature",
    volume = "248",
    pages = "30--31",
    year = "1974"
}

@article{Parikh:2004ih,
    author = "Parikh, Maulik K.",
    title = "{A Secret tunnel through the horizon}",
    eprint = "hep-th/0405160",
    archivePrefix = "arXiv",
    reportNumber = "CU-TP-1114",
    doi = "10.1142/S0218271804006498",
    journal = "Int. J. Mod. Phys. D",
    volume = "13",
    pages = "2351--2354",
    year = "2004"
}

@article{Parikh:1999mf,
    author = "Parikh, Maulik K. and Wilczek, Frank",
    title = "{Hawking radiation as tunneling}",
    eprint = "hep-th/9907001",
    archivePrefix = "arXiv",
    reportNumber = "PUPT-1775, SPIN-1998-12, IASSNS-HEP-98-22",
    doi = "10.1103/PhysRevLett.85.5042",
    journal = "Phys. Rev. Lett.",
    volume = "85",
    pages = "5042--5045",
    year = "2000"
}

@article{Hawking:1975vcx,
    author = "Hawking, S. W.",
    editor = "Gibbons, G. W. and Hawking, S. W.",
    title = "{Particle Creation by Black Holes}",
    doi = "10.1007/BF02345020",
    journal = "Commun. Math. Phys.",
    volume = "43",
    pages = "199--220",
    year = "1975",
    note = "[Erratum: Commun.Math.Phys. 46, 206 (1976)]"
}

@article{Peltola:2008jx,
    author = "Peltola, Ari",
    title = "{Local Approach to Hawking Radiation}",
    eprint = "0807.3309",
    archivePrefix = "arXiv",
    primaryClass = "gr-qc",
    doi = "10.1088/0264-9381/26/3/035014",
    journal = "Class. Quant. Grav.",
    volume = "26",
    pages = "035014",
    year = "2009"
}

@article{PADMANABHAN200549,
title = {Gravity and the thermodynamics of horizons},
journal = {Physics Reports},
volume = {406},
number = {2},
pages = {49-125},
year = {2005},
issn = {0370-1573},
doi = {https://doi.org/10.1016/j.physrep.2004.10.003},
url = {https://www.sciencedirect.com/science/article/pii/S0370157304004582},
author = {T. Padmanabhan}
}

@article{Zhang:2022vsl,
    author = "Zhang, Cong and Liu, Hongguang and Han, Muxin",
    title = "{Fermions in loop quantum gravity and resolution of doubling problem}",
    eprint = "2212.00933",
    archivePrefix = "arXiv",
    primaryClass = "gr-qc",
    doi = "10.1088/1361-6382/acf26b",
    journal = "Class. Quant. Grav.",
    volume = "40",
    number = "20",
    pages = "205022",
    year = "2023"
}

@article{Long_2025,
doi = {10.1088/1361-6382/adcb14},
url = {https://doi.org/10.1088/1361-6382/adcb14},
year = {2025},
month = {apr},
publisher = {IOP Publishing},
volume = {42},
number = {9},
pages = {095004},
author = {Long, Gaoping},
title = {Twisted geometric parametrization of holonomy-flux phase space in all dimensional loop quantum gravity},
journal = {Classical and Quantum Gravity}
}

@article{Han:2024rqb,
   title = {Spin foam amplitude of the black-to-white hole transition},
  author = {Han, Muxin and Qu, Dongxue and Zhang, Cong},
  journal = {Phys. Rev. D},
  volume = {110},
  issue = {12},
  pages = {124055},
  numpages = {34},
  year = {2024},
  month = {Dec},
  publisher = {American Physical Society},
  doi = {10.1103/PhysRevD.110.124055},
  url = {https://link.aps.org/doi/10.1103/PhysRevD.110.124055},
    eprint = "2404.02796",
    archivePrefix = "arXiv",
    primaryClass = "gr-qc"
}

@article{Han:2024ydv,
  title = {Cosmological dynamics from covariant loop quantum gravity with scalar matter},
  author = {Han, Muxin and Liu, Hongguang and Qu, Dongxue and Vidotto, Francesca and Zhang, Cong},
  journal = {Phys. Rev. D},
  volume = {111},
  issue = {8},
  pages = {086012},
  numpages = {22},
  year = {2025},
  month = {Apr},
  publisher = {American Physical Society},
  doi = {10.1103/PhysRevD.111.086012},
  url = {https://link.aps.org/doi/10.1103/PhysRevD.111.086012},
    eprint = "2402.07984",
    archivePrefix = "arXiv",
    primaryClass = "gr-qc"
}

@article{Bojowald:2001xe,
	Archiveprefix = {arXiv},
	Author = {Bojowald, Martin},
	Doi = {10.1103/PhysRevLett.86.5227},
	Eprint = {gr-qc/0102069},
	Journal = {Phys. Rev. Lett.},
	Pages = {5227-5230},
	Primaryclass = {gr-qc},
	Reportnumber = {CGPG-01-2-1},
	Slaccitation = {%%CITATION = GR-QC/0102069;%%},
	Title = {{Absence of singularity in loop quantum cosmology}},
	Volume = {86},
	Year = {2001}}

@article{Zhang:2024khj,
    author = "Zhang, Cong and Lewandowski, Jerzy and Ma, Yongge and Yang, Jinsong",
    title = "{Black holes and covariance in effective quantum gravity}",
    eprint = "2407.10168",
    archivePrefix = "arXiv",
    primaryClass = "gr-qc",
    doi = "10.1103/PhysRevD.111.L081504",
    journal = "Phys. Rev. D",
    volume = "111",
    number = "8",
    pages = "L081504",
    year = "2025"
}

@article{Zhang:2024ney,
    author = "Zhang, Cong and Lewandowski, Jerzy and Ma, Yongge and Yang, Jinsong",
    title = "{Black holes and covariance in effective quantum gravity: A solution without Cauchy horizons}",
    eprint = "2412.02487",
    archivePrefix = "arXiv",
    primaryClass = "gr-qc",
    doi = "10.1103/d6ks-d576",
    journal = "Phys. Rev. D",
    volume = "112",
    number = "4",
    pages = "044054",
    year = "2025"
}

@Inbook{Gambini2023,
author="Gambini, Rodolfo
and Olmedo, Javier
and Pullin, Jorge",
editor="Bambi, Cosimo
and Modesto, Leonardo
and Shapiro, Ilya",
title="Quantum Geometry and Black Holes",
bookTitle="Handbook of Quantum Gravity",
year="2023",
publisher="Springer Nature Singapore",
address="Singapore",
pages="1--34",
isbn="978-981-19-3079-9",
doi="10.1007/978-981-19-3079-9_105-1",
url="https://doi.org/10.1007/978-981-19-3079-9_105-1"
}

@article{Bojowald:2005cb,
    author = "Bojowald, Martin and Swiderski, Rafal",
    title = "{Spherically symmetric quantum geometry: Hamiltonian constraint}",
    eprint = "gr-qc/0511108",
    archivePrefix = "arXiv",
    reportNumber = "AEI-2005-171, NI05065",
    doi = "10.1088/0264-9381/23/6/015",
    journal = "Class. Quant. Grav.",
    volume = "23",
    pages = "2129--2154",
    year = "2006"
}

@article{Ashtekar:2005qt,
    author = "Ashtekar, Abhay and Bojowald, Martin",
    title = "{Quantum geometry and the Schwarzschild singularity}",
    eprint = "gr-qc/0509075",
    archivePrefix = "arXiv",
    reportNumber = "IGPG-05-09-01, AEI-2005-132",
    doi = "10.1088/0264-9381/23/2/008",
    journal = "Class. Quant. Grav.",
    volume = "23",
    pages = "391--411",
    year = "2006"
}

@article{Corichi:2015xia,
    author = "Corichi, Alejandro and Singh, Parampreet",
    title = "{Loop quantization of the Schwarzschild interior revisited}",
    eprint = "1506.08015",
    archivePrefix = "arXiv",
    primaryClass = "gr-qc",
    doi = "10.1088/0264-9381/33/5/055006",
    journal = "Class. Quant. Grav.",
    volume = "33",
    number = "5",
    pages = "055006",
    year = "2016"
}

@article{Han:2022rsx,
    title = {Covariant $\overline{\ensuremath{\mu}}$-scheme effective dynamics, mimetic gravity, and nonsingular black holes: Applications to spherically symmetric quantum gravity},
  author = {Han, Muxin and Liu, Hongguang},
  journal = {Phys. Rev. D},
  volume = {109},
  issue = {8},
  pages = {084033},
  numpages = {29},
  year = {2024},
  month = {Apr},
  publisher = {American Physical Society},
  doi = {10.1103/PhysRevD.109.084033},
  url = {https://link.aps.org/doi/10.1103/PhysRevD.109.084033},
    eprint = "2212.04605",
    archivePrefix = "arXiv",
    primaryClass = "gr-qc"
}

@article{PhysRevLett.102.051301,
  title = {Effective Scenario of Loop Quantum Cosmology},
  author = {Ding, You and Ma, Yongge and Yang, Jinsong},
  journal = {Phys. Rev. Lett.},
  volume = {102},
  issue = {5},
  pages = {051301},
  numpages = {4},
  year = {2009},
  month = {Feb},
  publisher = {American Physical Society},
  doi = {10.1103/PhysRevLett.102.051301},
  url = {https://link.aps.org/doi/10.1103/PhysRevLett.102.051301}
}

@article{Giesel:2023hys,
    title = {Generalized analysis of a dust collapse in effective loop quantum gravity: Fate of shocks and covariance},
  author = {Giesel, Kristina and Liu, Hongguang and Singh, Parampreet and Weigl, Stefan Andreas},
  journal = {Phys. Rev. D},
  volume = {110},
  issue = {10},
  pages = {104016},
  numpages = {23},
  year = {2024},
  month = {Nov},
  publisher = {American Physical Society},
  doi = {10.1103/PhysRevD.110.104016},
  url = {https://link.aps.org/doi/10.1103/PhysRevD.110.104016},
    eprint = "2308.10953",
    archivePrefix = "arXiv",
    primaryClass = "gr-qc",
}

@article{Han:2020uhb,
    author = "Han, Muxin and Liu, Hongguang",
    title = "{Improved effective dynamics of loop-quantum-gravity black hole and Nariai limit}",
    eprint = "2012.05729",
    archivePrefix = "arXiv",
    primaryClass = "gr-qc",
    doi = "10.1088/1361-6382/ac44a0",
    journal = "Class. Quant. Grav.",
    volume = "39",
    number = "3",
    pages = "035011",
    year = "2022"
}

@article{Ashtekar:2018lag,
    author = "Ashtekar, Abhay and Olmedo, Javier and Singh, Parampreet",
    title = "{Quantum Transfiguration of Kruskal Black Holes}",
    eprint = "1806.00648",
    archivePrefix = "arXiv",
    primaryClass = "gr-qc",
    doi = "10.1103/PhysRevLett.121.241301",
    journal = "Phys. Rev. Lett.",
    volume = "121",
    number = "24",
    pages = "241301",
    year = "2018"
}

@article{Olmedo:2017lvt,
	Archiveprefix = {arXiv},
	Author = {Olmedo, Javier and Saini, Sahil and Singh, Parampreet},
	Doi = {10.1088/1361-6382/aa8da8},
	Eprint = {1707.07333},
	Journal = {Class. Quant. Grav.},
	Number = {22},
	Pages = {225011},
	Primaryclass = {gr-qc},
	Slaccitation = {%%CITATION = ARXIV:1707.07333;%%},
	Title = {{From black holes to white holes: a quantum gravitational, symmetric bounce}},
	Volume = {34},
	Year = {2017}}

@article{Dadhich:2015ora,
    author = "Dadhich, Naresh and Joe, Anton and Singh, Parampreet",
    title = "{Emergence of the product of constant curvature spaces in loop quantum cosmology}",
    eprint = "1505.05727",
    archivePrefix = "arXiv",
    primaryClass = "gr-qc",
    doi = "10.1088/0264-9381/32/18/185006",
    journal = "Class. Quant. Grav.",
    volume = "32",
    number = "18",
    pages = "185006",
    year = "2015"
}

@article{Kelly:2020lec,
    author = "Kelly, Jarod George and Santacruz, Robert and Wilson-Ewing, Edward",
    title = "{Black hole collapse and bounce in effective loop quantum gravity}",
    eprint = "2006.09325",
    archivePrefix = "arXiv",
    primaryClass = "gr-qc",
    doi = "10.1088/1361-6382/abd3e2",
    journal = "Class. Quant. Grav.",
    volume = "38",
    number = "4",
    pages = "04LT01",
    year = "2021"
}

@article{Gambini:2013hna,
    author = "Gambini, Rodolfo and Olmedo, Javier and Pullin, Jorge",
    title = "{Quantum black holes in Loop Quantum Gravity}",
    eprint = "1310.5996",
    archivePrefix = "arXiv",
    primaryClass = "gr-qc",
    doi = "10.1088/0264-9381/31/9/095009",
    journal = "Class. Quant. Grav.",
    volume = "31",
    pages = "095009",
    year = "2014"
}

@article{Ashtekar:2006wn,
    author = "Ashtekar, Abhay and Pawlowski, Tomasz and Singh, Parampreet",
    title = "{Quantum Nature of the Big Bang: Improved dynamics}",
    eprint = "gr-qc/0607039",
    archivePrefix = "arXiv",
    reportNumber = "IGPG-06-7-2",
    doi = "10.1103/PhysRevD.74.084003",
    journal = "Phys. Rev. D",
    volume = "74",
    pages = "084003",
    year = "2006"
}

@book{first30years,
title = "Loop quantum gravity: The first 30 years",
author = "Abhay Ashtekar and Jorge Pulliny",
year = "2017",
month = "mar",
day = "16",
doi = "10.1142/10445",
language = "English (US)",
isbn = "9789813209930",
publisher = "World Scientific Publishing Co. Pte Ltd",
address = "Singapore",
}

@article{Long:2024lbd,
    author = "Long, Gaoping and Chen, Qian and Yang, Jinsong",
    title = "{Entanglement entropy of coherent intertwiner in loop quantum gravity}",
    eprint = "2403.18020",
    archivePrefix = "arXiv",
    primaryClass = "gr-qc",
    doi = "10.1103/PhysRevD.110.064017",
    journal = "Phys. Rev. D",
    volume = "110",
    number = "6",
    pages = "064017",
    year = "2024"
}

@article{Basu:2009cw,
    author = "Basu, Rudranil and Kaul, Romesh K. and Majumdar, Parthasarathi",
    title = "{Entropy of Isolated Horizons revisited}",
    eprint = "0907.0846",
    archivePrefix = "arXiv",
    primaryClass = "gr-qc",
    doi = "10.1103/PhysRevD.82.024007",
    journal = "Phys. Rev. D",
    volume = "82",
    pages = "024007",
    year = "2010"
}

@article{Engle:2010kt,
    author = "Engle, Jonathan and Noui, Karim and Perez, Alejandro and Pranzetti, Daniele",
    title = "{Black hole entropy from an SU(2)-invariant formulation of Type I isolated horizons}",
    eprint = "1006.0634",
    archivePrefix = "arXiv",
    primaryClass = "gr-qc",
    doi = "10.1103/PhysRevD.82.044050",
    journal = "Phys. Rev. D",
    volume = "82",
    pages = "044050",
    year = "2010"
}

@article{Kaul:2000kf,
    author = "Kaul, Romesh K. and Majumdar, Parthasarathi",
    title = "{Logarithmic correction to the Bekenstein-Hawking entropy}",
    eprint = "gr-qc/0002040",
    archivePrefix = "arXiv",
    doi = "10.1103/PhysRevLett.84.5255",
    journal = "Phys. Rev. Lett.",
    volume = "84",
    pages = "5255--5257",
    year = "2000"
}

@article{Perez:2014ura,
    author = "Perez, Alejandro",
    title = "{Statistical and entanglement entropy for black holes in quantum geometry}",
    eprint = "1405.7287",
    archivePrefix = "arXiv",
    primaryClass = "gr-qc",
    doi = "10.1103/PhysRevD.90.084015",
    journal = "Phys. Rev. D",
    volume = "90",
    number = "8",
    pages = "084015",
    year = "2014",
    note = "[Addendum: Phys.Rev.D 90, 089907 (2014)]"
}

@article{Song:2022zit,
    author = "Song, Shupeng and Long, Gaoping and Zhang, Cong and Zhang, Xiangdong",
    title = "{Thermodynamics of isolated horizons in loop quantum gravity}",
    eprint = "2205.09984",
    archivePrefix = "arXiv",
    primaryClass = "gr-qc",
    doi = "10.1103/PhysRevD.106.126007",
    journal = "Phys. Rev. D",
    volume = "106",
    number = "12",
    pages = "126007",
    year = "2022"
}

@article{Ashtekar:2000eq,
    author = "Ashtekar, A. and Baez, John C. and Krasnov, Kirill",
    title = "{Quantum geometry of isolated horizons and black hole entropy}",
    eprint = "gr-qc/0005126",
    archivePrefix = "arXiv",
    reportNumber = "NSF-ITP-99-153",
    doi = "10.4310/ATMP.2000.v4.n1.a1",
    journal = "Adv. Theor. Math. Phys.",
    volume = "4",
    pages = "1--94",
    year = "2000"
}

@article{Arnowitt:1962hi,
    author = "Arnowitt, Richard L. and Deser, Stanley and Misner, Charles W.",
    title = "{The Dynamics of general relativity}",
    eprint = "gr-qc/0405109",
    archivePrefix = "arXiv",
    doi = "10.1007/s10714-008-0661-1",
    journal = "Gen. Rel. Grav.",
    volume = "40",
    pages = "1997--2027",
    year = "2008"
}

@article{Arnowitt:1960es,
    author = "Arnowitt, Richard L. and Deser, Stanley and Misner, Charles W.",
    title = "{Canonical variables for general relativity}",
    doi = "10.1103/PhysRev.117.1595",
    journal = "Phys. Rev.",
    volume = "117",
    pages = "1595--1602",
    year = "1960"
}

@article{Ashtekar:1997yu,
    author = "Ashtekar, A. and Baez, J. and Corichi, A. and Krasnov, Kirill",
    title = "{Quantum geometry and black hole entropy}",
    eprint = "gr-qc/9710007",
    archivePrefix = "arXiv",
    reportNumber = "CGPG-97-9-3",
    doi = "10.1103/PhysRevLett.80.904",
    journal = "Phys. Rev. Lett.",
    volume = "80",
    pages = "904--907",
    year = "1998"
}

@article{Song:2020arr,
    author = "Song, Shupeng and Li, Haida and Ma, Yongge and Zhang, Cong",
    title = "{Entropy of black holes with arbitrary shapes in loop quantum gravity}",
    eprint = "2002.08869",
    archivePrefix = "arXiv",
    primaryClass = "gr-qc",
    doi = "10.1007/s11433-021-1770-3",
    journal = "Sci. China Phys. Mech. Astron.",
    volume = "64",
    number = "12",
    pages = "120411",
    year = "2021"
}

@article{Ghosh:2011fc,
    author = "Ghosh, Amit and Perez, Alejandro",
    title = "{Black hole entropy and isolated horizons thermodynamics}",
    eprint = "1107.1320",
    archivePrefix = "arXiv",
    primaryClass = "gr-qc",
    doi = "10.1103/PhysRevLett.107.241301",
    journal = "Phys. Rev. Lett.",
    volume = "107",
    pages = "241301",
    year = "2011",
    note = "[Erratum: Phys.Rev.Lett. 108, 169901 (2012)]"
}

@article{Long0propagator,
  title = {Semiclassical propagator for coherent state on twisted geometry},
  author = {Long, Gaoping and Liu, Hongguang and Zhang, Cong},
  journal = {Phys. Rev. D},
  volume = {111},
  issue = {4},
  pages = {046021},
  numpages = {21},
  year = {2025},
  month = {Feb},
  publisher = {American Physical Society},
  doi = {10.1103/PhysRevD.111.046021},
  url = {https://link.aps.org/doi/10.1103/PhysRevD.111.046021}
}

@article{Longrepresentationtwisted,
  title = {Quantum representation of reduced twisted geometry in loop quantum gravity},
  author = {Long, Gaoping and Zhang, Cong and Liu, Hongguang},
  journal = {Phys. Rev. D},
  volume = {112},
  issue = {2},
  pages = {024022},
  numpages = {21},
  year = {2025},
  month = {Jul},
  publisher = {American Physical Society},
  doi = {10.1103/99fq-xz2w},
  url = {https://link.aps.org/doi/10.1103/99fq-xz2w}
}

@article{Long:2020agv,
    author = "Long, Gaoping and Ma, Yongge",
    title = "{Polytopes in all dimensional loop quantum gravity}",
    journal = {Eur. Phys. J. C},
  volume = {82},
  number = {41},
  year = {2022},
  doi = {10.1140/epjc/s10052-022-09988-2}
}

@article{Bianchi:2008es,
    author = "Bianchi, Eugenio",
    title = "{The Length operator in Loop Quantum Gravity}",
    eprint = "0806.4710",
    archivePrefix = "arXiv",
    primaryClass = "gr-qc",
    doi = "10.1016/j.nuclphysb.2008.08.013",
    journal = "Nucl. Phys. B",
    volume = "807",
    pages = "591--624",
    year = "2009"
}

@article{PhysRevD.83.044035,
  title = {Polyhedra in loop quantum gravity},
  author = {Bianchi, Eugenio and Don\'a, Pietro and Speziale, Simone},
  journal = {Phys. Rev. D},
  volume = {83},
  issue = {4},
  pages = {044035},
  numpages = {17},
  year = {2011},
  month = {Feb},
  publisher = {American Physical Society},
  doi = {10.1103/PhysRevD.83.044035},
  url = {https://link.aps.org/doi/10.1103/PhysRevD.83.044035}
}

@article{Long:2020euh,
    author = "Long, Gaoping and Bodendorfer, Norbert",
    title = "{Perelomov-type coherent states of SO($D+1$) in all-dimensional loop quantum gravity}",
    eprint = "2006.13122",
    archivePrefix = "arXiv",
    primaryClass = "gr-qc",
    doi = "10.1103/PhysRevD.102.126004",
    journal = "Phys. Rev. D",
    volume = "102",
    number = "12",
    pages = "126004",
    year = "2020"
}

@article{PhysRevD.104.046014,
  title = {Superposition type coherent states in all dimensional loop quantum gravity},
  author = {Long, Gaoping and Zhang, Cong and Zhang, Xiangdong},
  journal = {Phys. Rev. D},
  volume = {104},
  issue = {4},
  pages = {046014},
  numpages = {13},
  year = {2021},
  month = {Aug},
  publisher = {American Physical Society},
  doi = {10.1103/PhysRevD.104.046014},
  url = {https://link.aps.org/doi/10.1103/PhysRevD.104.046014}
}

@article{PhysRevD.103.086016,
  title = {Geometric parametrization of $\textsc{SO(D+1)}$ phase space of all dimensional loop quantum gravity},
  author = {Long, Gaoping and Lin, Chun-Yen},
  journal = {Phys. Rev. D},
  volume = {103},
  issue = {8},
  pages = {086016},
  numpages = {18},
  year = {2021},
  month = {Apr},
  publisher = {American Physical Society},
  doi = {10.1103/PhysRevD.103.086016},
  url = {https://link.aps.org/doi/10.1103/PhysRevD.103.086016}
}

@article{QoperatorPhysRevD.62.104021,
  title = {$\mathrm{Q\ifmmode \hat{}\else \^{}\fi{}}$ operator for canonical quantum gravity},
  author = {Ma, Yongge and Ling, Yi},
  journal = {Phys. Rev. D},
  volume = {62},
  issue = {10},
  pages = {104021},
  numpages = {6},
  year = {2000},
  month = {Oct},
  publisher = {American Physical Society},
  doi = {10.1103/PhysRevD.62.104021},
  url = {https://link.aps.org/doi/10.1103/PhysRevD.62.104021}
}

@article{volumePhysRevD.94.044003,
  title = {New volume and inverse volume operators for loop quantum gravity},
  author = {Yang, Jinsong and Ma, Yongge},
  journal = {Phys. Rev. D},
  volume = {94},
  issue = {4},
  pages = {044003},
  numpages = {16},
  year = {2016},
  month = {Aug},
  publisher = {American Physical Society},
  doi = {10.1103/PhysRevD.94.044003},
  url = {https://link.aps.org/doi/10.1103/PhysRevD.94.044003}
}

@article{Yang_2019Consistencycheck,
doi = {10.1088/1674-1137/43/10/103106},
url = {https://dx.doi.org/10.1088/1674-1137/43/10/103106},
year = {2019},
month = {oct},
publisher = {Chinese Physical Society and the Institute of High Energy Physics of the Chinese Academy of Sciences and the Institute of Modern Physics of the Chinese Academy of Sciences and IOP Publishing Ltd},
volume = {43},
number = {10},
pages = {103106},
author = {Jinsong Yang and Yongge Ma},
title = {Consistency check on the fundamental and alternative flux operators in loop quantum gravity *},
journal = {Chinese Physics C}
}

@article{Giesel_2006Consistencycheck,
doi = {10.1088/0264-9381/23/18/011},
url = {https://dx.doi.org/10.1088/0264-9381/23/18/011},
year = {2006},
month = {aug},
publisher = {},
volume = {23},
number = {18},
pages = {5667},
author = {K Giesel and T Thiemann},
title = {Consistency check on volume and triad operator quantization in loop quantum gravity: I},
journal = {Classical and Quantum Gravity}
}

@article{Ma:2010fy,
    author = "Ma, Yongge and Soo, Chopin and Yang, Jinsong",
    title = "{New length operator for loop quantum gravity}",
    eprint = "1004.1063",
    archivePrefix = "arXiv",
    primaryClass = "gr-qc",
    doi = "10.1103/PhysRevD.81.124026",
    journal = "Phys. Rev. D",
    volume = "81",
    pages = "124026",
    year = "2010"
}

@article{Calcinari_2020,
   title={Twisted geometries coherent states for loop quantum gravity},
   volume={38},
   ISSN={1361-6382},
   url={http://dx.doi.org/10.1088/1361-6382/abc273},
   DOI={10.1088/1361-6382/abc273},
   number={2},
   journal={Classical and Quantum Gravity},
   publisher={IOP Publishing},
   author={Calcinari, Andrea and Freidel, Laurent and Livine, Etera and Speziale, Simone},
   year={2020},
   month={Dec},
   pages={025004}
}

@article{Thiemann:2000bx,
    author = "Thiemann, T. and Winkler, O.",
    title = "{Gauge field theory coherent states (GCS): III. Ehrenfest theorems}",
    eprint = "hep-th/0005234",
    archivePrefix = "arXiv",
    reportNumber = "AEI-2000-029",
    doi = "10.1088/0264-9381/18/21/315",
    journal = "Class. Quant. Grav.",
    volume = "18",
    pages = "4629--4682",
    year = "2001"
}

@article{long2019coherent,
  title={Coherent intertwiner solution of simplicity constraint in all dimensional loop quantum gravity},
  author={Long, Gaoping and Lin, Chun-Yen and Ma, Yongge},
  journal={Physical Review D},
  volume={100},
  number={6},
  pages={064065},
  year={2019},
  publisher={APS}
}

@article{long2020operators,
    author = "Long, Gaoping and Ma, Yongge",
    title = "{General geometric operators in all dimensional loop quantum gravity}",
    eprint = "2003.03952",
    archivePrefix = "arXiv",
    primaryClass = "gr-qc",
    doi = "10.1103/PhysRevD.101.084032",
    journal = "Phys. Rev. D",
    volume = "101",
    number = "8",
    pages = "084032",
    year = "2020"
}

@article{Han2005FUNDAMENTAL,
  title={FUNDAMENTAL STRUCTURE OF LOOP QUANTUM GRAVITY},
  author={Han, Muxin and Yongge, M. A. and Huang, Weiming},
  journal={International Journal of Modern Physics D},
  volume={16},
  number={09},
  pages={1397-1474},
  year={2005},
}

@article{Zhang:2021qul,
    author = "Zhang, Cong and Song, Shicong and Han, Muxin",
    title = "{First-Order Quantum Correction in Coherent State Expectation Value of Loop-Quantum-Gravity Hamiltonian}",
    eprint = "2102.03591",
    archivePrefix = "arXiv",
    primaryClass = "gr-qc",
    doi = "10.1103/PhysRevD.105.064008",
    journal = "Phys. Rev. D",
    volume = "105",
    pages = "064008",
    year = "2022"
}

@article{Han:2019vpw,
      author         = "Han, Muxin and Liu, Hongguang",
      title          = "{Effective Dynamics from Coherent State Path Integral of
                        Full Loop Quantum Gravity}",
      journal        = "Phys. Rev.",
      volume         = "D101",
      year           = "2020",
      number         = "4",
      pages          = "046003",
      doi            = "10.1103/PhysRevD.101.046003",
      eprint         = "1910.03763",
      archivePrefix  = "arXiv",
      primaryClass   = "gr-qc",
      SLACcitation   = "%%CITATION = ARXIV:1910.03763;%%"
}

@article{Long:2021izw,
    author = "Long, Gaoping and Ma, Yongge",
    title = "{Effective dynamics of weak coupling loop quantum gravity}",
    eprint = "2111.11844",
    archivePrefix = "arXiv",
    primaryClass = "gr-qc",
    doi = "10.1103/PhysRevD.105.044043",
    journal = "Phys. Rev. D",
    volume = "105",
    number = "4",
    pages = "044043",
    year = "2022"
}

@article{LONG2025139580,
title = {Holonomy operator for spin connection in twisted geometry},
journal = {Physics Letters B},
volume = {866},
pages = {139580},
year = {2025},
issn = {0370-2693},
doi = {https://doi.org/10.1016/j.physletb.2025.139580},
url = {https://www.sciencedirect.com/science/article/pii/S0370269325003417},
author = {Gaoping Long and Hongguang Liu}
}

@article{Ashtekar2012Background,
  title={Background independent quantum gravity: a status report},
  author={Ashtekar, Abhay and Lewandowski, Jerzy},
  journal={Classical and Quantum Gravity},
  volume={21},
  number={15},
  pages={R53-R152},
  year={2012},
}

@article{Long:2022cex,
    author = "Long, Gaoping",
    title = "{Twisted geometry coherent states in all dimensional loop quantum gravity. II. Ehrenfest property}",
    eprint = "2204.03056",
    archivePrefix = "arXiv",
    primaryClass = "gr-qc",
    doi = "10.1103/PhysRevD.106.066021",
    journal = "Phys. Rev. D",
    volume = "106",
    number = "6",
    pages = "066021",
    year = "2022"
}

@article{Ashtekar:2011ni,
    author = "Ashtekar, Abhay and Singh, Parampreet",
    title = "{Loop Quantum Cosmology: A Status Report}",
    eprint = "1108.0893",
    archivePrefix = "arXiv",
    primaryClass = "gr-qc",
    doi = "10.1088/0264-9381/28/21/213001",
    journal = "Class. Quant. Grav.",
    volume = "28",
    pages = "213001",
    year = "2011"
}

@article{Long:2021lmd,
    author = "Long, Gaoping and Zhang, Xiangdong and Zhang, Cong",
    title = "{Twisted geometry coherent states in all dimensional loop quantum gravity: Construction and peakedness properties}",
    eprint = "2110.01317",
    archivePrefix = "arXiv",
    primaryClass = "gr-qc",
    doi = "10.1103/PhysRevD.105.066021",
    journal = "Phys. Rev. D",
    volume = "105",
    number = "6",
    pages = "066021",
    year = "2022"
}

@article{Long:2020oma,
    author = "Long, Gaoping and Liu, Yunlong and Zhang, Xiangdong",
    title = "{Energy conditions in the new model of loop quantum cosmology}",
    eprint = "2011.07712",
    archivePrefix = "arXiv",
    primaryClass = "gr-qc",
    doi = "10.1088/1674-1137/ac1e83",
    journal = "Chin. Phys. C",
    volume = "45",
    number = "11",
    pages = "115102",
    year = "2021"
}

@article{Zhang:2021zfp,
    author = "Zhang, Xiangdong and Long, Gaoping and Ma, Yongge",
    title = "{Loop quantum gravity and cosmological constant}",
    eprint = "2101.07527",
    archivePrefix = "arXiv",
    primaryClass = "gr-qc",
    doi = "10.1016/j.physletb.2021.136770",
    journal = "Phys. Lett. B",
    volume = "823",
    pages = "136770",
    year = "2021"
}

\appendix
\section{Quantization of gravitational field and the semiclassical spacetime geometry}\label{app1}
With gauge fixing \eqref{gauge1} and\eqref{gauge2},  the  dynamical variables remaining for the geometry sector  contain only the canonical pair $(K_\varphi,E^\varphi)$.  The loop quantization of the  reduced phase space $\bar{\mathcal{P}}^{\text{g}}$ coordinatized by  $(K_\varphi,E^\varphi)$  gives the kinematic Hilbert space  of quantum gravity in the spherical model \cite{Gambini:2013hna}. This space consists of  colored network states that encode  quantum geometry, represented by cylindrical functions defined on some graphs $\gamma$. 
For a given  graph $\gamma$ on the manifold $\mathbb{X}$, we associate the variable \(K_x\) with each elementary edge and the scalar \(K_{\varphi}\) with each vertex. Then, the Hilbert space  $\mathcal{H}_{\text{g}}^\gamma$ associated with $\gamma$ is spanned by  colored network states on $\gamma$, which take the form
\begin{equation}
    T_{\gamma, \vec{\mu}}(K_{\varphi}) = \prod_{v\in \gamma} \exp\left(\mathbf{i} \mu_v K_{\varphi}(v)\right),
\end{equation}
where   \(\mu_v \in \mathbb{R}\) is the color of the vertex \(v\),  and $\vec{\mu}=(\dots, \mu_{v_{\imath-1}}, \mu_{v_{\imath}},\mu_{v_{\imath+1}},\dots)$ denote all the colors associated with the edges and vertices  of the graph. 
The inner product among  colored network states on $\gamma$ is given by 
\begin{equation}
\langle \gamma, \vec{\mu} | \gamma,  \vec{\mu}'\rangle =  \delta_{\vec{\mu}, \vec{\mu}'} ,
\end{equation}
where  $| \gamma, \vec{\mu}\rangle$ represents the state function $   T_{\gamma, \vec{\mu}}$.
The fundamental operators in  $\mathcal{H}_{\text{g}}^\gamma$ associated with $E^\varphi_v :=\lim_{\epsilon\to 0}\int_{x(v)-\epsilon}^{x(v)+\epsilon}  dx E^\varphi$ are given by
\begin{equation}
\hat{E}^\varphi_v| \gamma, \vec{\mu}\rangle = G\hbar \mu_v | \gamma,  \vec{\mu}\rangle.
\end{equation}
 The fundamental operators in  $\mathcal{H}_{\text{g}}^\gamma$ associated with $   N^{\pm}_{v,\rho}:=\exp(\pm\mathbf{i}\rho K_\varphi(v)), 0<\rho\in\mathbb{R}$ are given by
\begin{equation}
  \hat{N}^{\pm}_{v,\rho}| \gamma,  \vec{\mu}\rangle =  | \gamma, \vec{\mu}_{v,\pm\rho}\rangle,
\end{equation}
where $\vec{\mu}_{v_\imath,\pm\rho}:=(\dots, \mu_{v_{\imath-1}}, \mu_{v_{\imath}}\pm \rho,\mu_{v_{\imath+1}},\dots)$.

The semiclassical geometry in this quantum theory is described by the coherent state. 
For an arbitrary given $\vec{\rho}=(\dots, \rho_{v_{\imath-1}}, \rho_{v_{\imath}},\rho_{v_{\imath+1}},\dots)$ and $\vec{\beta}=(\dots, \beta_{v_{\imath-1}}, \beta_{v_{\imath}},\beta_{v_{\imath+1}},\dots)$  with $0\leq\beta_v<\rho_v$, we have the subspace $\mathcal{H}^{\vec{\beta}}_\gamma\subset\mathcal{H}_{\text{g}}^\gamma$ defined by  $\mathcal{H}^{\vec{\beta}}_\gamma:=\{| \gamma, \vec{j}, \vec{\mu}=\vec{\beta}\pm \vec{n}\cdot \vec{\rho}\rangle|\  \forall \  \vec{n}= (\dots, n_{v_{\imath-1}}, n_{v_{\imath}},n_{v_{\imath+1}},\dots), \ n_v\in \mathbb{N}\}$.  The  coherent state in $\mathcal{H}^{\vec{\beta}}_\gamma$ can be established as \cite{Long:2024lbd,Long:2020euh,PhysRevD.104.046014,Calcinari_2020,
Thiemann:2000bx,
long2019coherent,Zhang:2021qul,
Han:2019vpw, Long:2022cex,Long:2021lmd} 
\begin{equation}
    T_{\gamma,  \vec{n}^0,\vec{\xi}^0}(K_{\varphi}) = \prod_{v\in \gamma} \sum_{n_v}\exp(\frac{-(n_v-n^0_v)^2}{t/\rho_v^2})\exp\left(\mathbf{i} (\beta_v+n_v \rho_v) (K_{\varphi}(v)-\xi_v^0)\right).
\end{equation}
Denoted by  $| \mathfrak{g}_\gamma\rangle\in \mathcal{H}^\gamma_{\text{g}}$ the normalized version of the coherent state $   T_{\gamma,  \vec{n}^0,\vec{\xi}^0}(K_{\varphi})$ with  
\begin{equation}
\mathfrak{g}_\gamma:=\{(E^{\varphi,0}_v, K^0_\varphi(v))|v\in \gamma\},
\end{equation}
where $ K^0_\varphi(v)=\xi^0_v$ and $E^{\varphi,0}_v=  (\beta_v+n^0_v \rho_v)G\hbar$. 
It can be verified the following three properties of  $|\mathfrak{g}_\gamma\rangle$. First, the overlap of  $|\mathfrak{g}_\gamma\rangle$ is estimated by
 
\begin{eqnarray}\label{overlap000}
\langle \mathfrak{g}_\gamma|\mathfrak{g}'_\gamma\rangle&\simeq&\prod_{v}  \exp\!\left(
 -\mathbf{i}\left( \beta_v + \frac{n_0(v) + n_0'(v)}{2} \, \rho \right)(\xi^0(v) - \xi'^0(v))
\right)\\\nonumber
&&\cdot  \exp\!\left(
 - \frac{\rho_v^2}{2t} (n_0(v) - n_0'(v))^2
 - \frac{t}{8} (\xi^0(v) - \xi'^0(v))^2
\right);
\end{eqnarray}
Second, the expectation values of the basic operators in  $|\mathfrak{g}_\gamma\rangle$ are given by
\begin{eqnarray}
\langle \mathfrak{g}_\gamma  |  \hat{N}^{\pm}_{v,\rho_v}|\mathfrak{g}_\gamma \rangle
\approx \exp(\pm\mathbf{i}\rho_v \xi^0_v),\quad  \langle \mathfrak{g}_\gamma  |\hat{E}^\varphi_v/(G\hbar)|\mathfrak{g}_\gamma \rangle
\approx\beta_v+n^0_v \rho_v
\end{eqnarray}
with the corresponding uncertainty being given by
\begin{eqnarray}
\Delta\langle \mathfrak{g}_\gamma  |  \hat{N}^{\pm}_{v,\rho_v}|\mathfrak{g}_\gamma \rangle
\propto \frac{|\rho_v|}{\sqrt{t}} ,\quad  \Delta\langle \mathfrak{g}_\gamma  |\hat{E}^\varphi_v/(G\hbar)|\mathfrak{g}_\gamma \rangle
\propto \sqrt{t};
\end{eqnarray}
Third,  $|\mathfrak{g}_\gamma\rangle$  satisfies the Ehrenfest property as
\begin{eqnarray}\label{gexpect000}
\langle \mathfrak{g}_\gamma  |\hat{O}_\gamma|\mathfrak{g}_\gamma' \rangle
=\langle \mathfrak{g}_\gamma  |\mathfrak{g}_\gamma' \rangle O_\gamma(\mathfrak{g}_\gamma') (1+\mathcal{O}(t)),
\end{eqnarray}
where $O_\gamma(\mathfrak{g}_\gamma')$ is a regular functional of $\{(E^{\varphi}_v,  N^{\pm}_{v,\rho_v})|{v\in\gamma}\}$, and $\hat{O}_\gamma$ is the operator corresponding to $O_\gamma(\mathfrak{g}_\gamma')$. These properties  of the coherent state  $|\mathfrak{g}_\gamma\rangle$ give a semiclassical and discrete description of the spacetime geometry. 

Let us consider the semiclassical and continuum description of the spacetime geometry. Denoted by $\mu_\gamma$ the coordinate distance of the adjacent vertices in $\gamma$.
We can consider the above construction in the limit  $\mu_{\gamma}\to0$, and we have 
\begin{eqnarray}
    \mathfrak{g} \underset{\mu_\gamma\to0}{=}\mathfrak{g}_\gamma,\quad    | \mathfrak{g}\rangle \underset{\mu_\gamma\to0}{:=}|\mathfrak{g}_\gamma \rangle.
\end{eqnarray}
Correspondingly, we have
\begin{eqnarray}\label{phasequan}
\bar{\mathcal{P}}^{\text{g},\text{qua}}\underset{\mu_\gamma\to0}{=}\mathcal{P}^{\text{g}}_\gamma,\quad \mathcal{H}_{\text{g}}\underset{\mu_\gamma\to0}{:=}\mathcal{H}^\gamma_{\text{g}} ;
\end{eqnarray}
and 
\begin{eqnarray}
   O \underset{\mu_\gamma\to0}{=}O_\gamma,\quad    \hat{O} \underset{\mu_\gamma\to0}{:=}\hat{O} _\gamma ,
\end{eqnarray}
where $\bar{\mathcal{P}}^{\text{g},\text{qua}}\ni\mathfrak{g}$ , $\mathcal{P}^{\text{g}}_\gamma\ni\mathfrak{g}_\gamma $,  $O$ is the regular functional on $\bar{\mathcal{P}}^{\text{g},\text{qua}}$, and  the phase space $\bar{\mathcal{P}}^{\text{g},\text{qua}}$  is a quantum generalization of $\bar{\mathcal{P}}^{\text{g}}$  with $\bar{\mathcal{P}}^{\text{g}}$ being a dense subspace of $\bar{\mathcal{P}}^{\text{g},\text{qua}}$. In particular,   the configurations of the phase space $\bar{\mathcal{P}}^{\text{g},\text{qua}}$  include the discontinuous geometric fields,  since these fields are given by a continuum limit of discrete geometries  \cite{LONG2025139580,Long:2020agv,Han2005FUNDAMENTAL,PhysRevD.83.044035}. Finally, in the limit $\mu_\gamma\to0$,   the overlap \eqref{overlap000} and the Ehrenfest property \eqref{gexpect000}   of  $|\mathfrak{g}_\gamma\rangle$ are given as
 \begin{eqnarray}
\langle \mathfrak{g}  |\mathfrak{g}' \rangle
\sim \exp(|\mathfrak{g}  -\mathfrak{g}'|^2/t)
\end{eqnarray}
and
\begin{eqnarray}
\langle \mathfrak{g}  |\hat{O}|\mathfrak{g}' \rangle
=  \langle \mathfrak{g}  |\mathfrak{g}' \rangle O(\mathfrak{g}')(1+\mathcal{O}(t)),
\end{eqnarray}
respectively, which are just the Eqs. \eqref {overlap222} and  \eqref{gexpect002}  used in section \ref{sec301}.

\section{An example of the unitary Bogoliubov transformation \eqref{Bogo}  }\label{app2}
Recall  the subspace $\mathcal{U}_{ v,v'}\subset \mathcal{S}_{ v,v'}$ defined by
\begin{eqnarray}
\mathcal{U}_{ v,v'}:=\{\mathfrak{g}''\in\mathcal{S}_{ v,v'}|\int dkdk'| \beta_{kk'}(\mathfrak{g}',\mathfrak{g})|^2< \infty, \  \forall \mathfrak{g}',\mathfrak{g}\in \mathcal{U}_{ v,v'} \}
\end{eqnarray}
and the Bogoliubov coefficients $\beta_{kk'}(\mathfrak{g}',\mathfrak{g})$ given by
 \begin{eqnarray}\label{}
\beta_{kk'}(\mathfrak{g}',\mathfrak{g})=\frac{1}{2}\sqrt{\frac{\omega_{k}}{\omega_{k'}}}I_{\mathfrak{g}}-\frac{1}{2}\sqrt{\frac{\omega_{k'}}{\omega_{k}}}I_{{\mathfrak{g}}'}=\frac{1}{2}I_{\mathfrak{g}}(\sqrt{\frac{\omega_{k}}{\omega_{k'}}}-\sqrt{\frac{\omega_{k'}}{\omega_{k}}})+\frac{1}{2}\sqrt{\frac{\omega_{k'}}{\omega_{k}}}(I_{{\mathfrak{g}}}-I_{{\mathfrak{g}}'})
\end{eqnarray}
with $I_{\mathfrak{g}}:=\lim_{L\to\infty}\int_{-L}^{+L} d\check{x}^\ast_{\mathfrak{g}} {e^{-\mathbf{i}k\check{x}^\ast_{\mathfrak{g}}}}{e^{-\mathbf{i}k'\check{x}^\ast_{\mathfrak{g}'}}}$ and $I_{\mathfrak{g}'}:=\lim_{L\to\infty}\int_{-L}^{+L} d\check{x}^\ast_{\mathfrak{g}'} {e^{-\mathbf{i}k\check{x}^\ast_{\mathfrak{g}}}}{e^{-\mathbf{i}k'\check{x}^\ast_{\mathfrak{g}'}}}$.
One can introduce a set of $\mathfrak{g}_{M}\in \mathcal{S}_{v,v'}$ parametrized by $M$ with $x(v)=2GM_v\leq 2G M<x(v')$;
Specifically, it is given by
\begin{equation}
(E^\varphi)^2|_{\mathfrak{g}_{M}}=\begin{cases}
 \frac{x^2}{1-2G M/x}, &x= x(v'),\\
  \frac{x^2}{1-2G M_{v}/x},     & x(v)<x<x(v').
\end{cases}
\end{equation}
Then, an example of the space $\mathcal{U}_{ v,v'}$ can be given as 
\begin{eqnarray}\label{exampleS}
\mathcal{U}_{ v,v'}\equiv\{\mathfrak{g}=\mathfrak{g}_{M}\in\mathcal{S}_{ v,v'}| x(v)\leq 2G M< x(v')\}.
\end{eqnarray}

It is necessary to verify whether $\int dkdj| \beta_{kj}(\mathfrak{g}',\mathfrak{g})|^2< \infty $ holds for arbitrary $\mathfrak{g}',\mathfrak{g}\in \mathcal{U}_{ v,v'}$.
Let us consider arbitrary $\mathfrak{g}'=\mathfrak{g}_{M'},\mathfrak{g}=\mathfrak{g}_{M}\in \mathcal{U}_{ v,v'}$ with $M<M'$, one has 
\begin{equation}
\frac{\partial {\check{x}^\ast_{\mathfrak{g}}}}{\partial {\check{x}^\ast_{\mathfrak{g}'}}}=\frac{(E^\varphi)^2|_{\mathfrak{g}}}{(E^\varphi)^2|_{\mathfrak{g}'}}=\begin{cases}
\text{bounded and positive real number}, &x=x(v'),\\
1,     & x(v)<x<x(v').
\end{cases}
\end{equation}
One can choose the  boundary condition that ensures $\check{x}^\ast_{\mathfrak{g}}=\check{x}^\ast_{\mathfrak{g}'}$ in $x(v)<x<x(v')$, and notice that the point $x=x(v')$ is located at $\check{x}^\ast_{\mathfrak{g}}\to +\infty $ in the limit $L\to \infty$. Then, it is direct to get 
\begin{eqnarray}
I_{\mathfrak{g}}=\lim_{L\to\infty}\int_{-L}^{+L} d\check{x}^\ast_{\mathfrak{g}} {e^{-\mathbf{i}k\check{x}^\ast_{\mathfrak{g}}}}{e^{-\mathbf{i}k'\check{x}^\ast_{\mathfrak{g}'}}}=2\pi \delta(k+k').
\end{eqnarray}
  Furthermore, it is direct to verify that the Bogoliubov coefficients $\beta_{kj}(\mathfrak{g}',\mathfrak{g})$  for our chosen $\mathfrak{g}'=\mathfrak{g}_{M'},\mathfrak{g}=\mathfrak{g}_{M}$ satisfy
\begin{eqnarray}
\int dkdk'| \beta_{kk'}(\mathfrak{g}',\mathfrak{g})|^2=0.
\end{eqnarray}
Now, we can conclude that $\int dkdj| \beta_{kj}(\mathfrak{g}',\mathfrak{g})|^2< \infty $ holds for arbitrary $\mathfrak{g}',\mathfrak{g}\in \mathcal{U}_{ v,v'}$ , so  $\mathcal{U}_{ v,v'}$ defined by \eqref{exampleS} is an example of the space $\mathcal{U}_{ v,v'}$ exactly.

It is necessary to emphasize that  the space $\mathcal{U}_{ v,v'}$ given by \eqref{exampleS} contains the geometry fields which are discontinuous at $v'$. It seems that the space $\mathcal{U}_{ v,v'}$  is not a subspace of  the classical reduced phase space $\bar{\mathcal{P}}^{\text{g}}$, since  $\bar{\mathcal{P}}^{\text{g}}$ contains no discontinuous geometry field.  In fact, the loop representation  extends our consideration to the enlarged  phase space $\bar{\mathcal{P}}^{\text{g},\text{qua}}$ defined by \eqref{phasequan}, which contains  discontinuous geometry field configuration  \cite{LONG2025139580,Han2005FUNDAMENTAL}. In particular, the space $\mathcal{U}_{ v,v'}$ given by \eqref{exampleS} is a subspace of $\bar{\mathcal{P}}^{\text{g},\text{qua}}$ exactly.

\section{Local Hartle-Hawking vacuum on classical geometry }\label{app3}
Consider the spherical spacetime metric generated by the phase space configuration $\mathfrak{g}\in\mathcal{S}_{ v,v'}$ , which reads  
\begin{equation}\label{metric000}
ds^2=-N^2 dt^2+\frac{(E^\varphi)^2}{E^x}dx^2+E^x (d\theta^2+\sin^2\theta d\varphi^2).
\end{equation}
One can fix the spacetime coordinate by $N=\frac{\sqrt{E^x}}{E^\varphi}$ and $E^x=x^2$.  Then, we denote $h_{\mathfrak{g}}(x)=(E^\varphi)^2$ and  the metric \eqref{metric000} can be given as
\begin{equation}\label{metric0001}
ds^2=-\frac{x^2}{h_{\mathfrak{g}}(x)}dt^2+\frac{h_{\mathfrak{g}}(x)}{x^2}dx^2+x^2(d\theta^2+\sin^2\theta d\varphi^2),
\end{equation}
 We introduce the tortoise coordinate $x^\ast_{\mathfrak{g}}$ by the transformation  
\begin{equation}
\frac{\partial x}{\partial {x^\ast_{\mathfrak{g}}}}=\frac{x^2}{h_{\mathfrak{g}}(x)},
\end{equation}
 with  the tortoise coordinate $x^\ast_{\mathfrak{g}}$   satisfying
\begin{equation}
\lim_{x\to x_{\text{h}}} x^\ast_{\mathfrak{g}}\to-\infty, \quad \lim_{x\to x_{\text{h}}+\delta} x^\ast_{\mathfrak{g}}\to\delta^\ast_{\mathfrak{g}},\quad 0<\delta^\ast_{\mathfrak{g}}<+\infty.
\end{equation}
In the tortoise coordinate $x^\ast_{\mathfrak{g}}$, the  metric in the region $x_{\text{h}}< x\leq x_{\text{h}}+\delta$ can be written as
\begin{equation}\label{metric2}
ds^2=-\frac{x^2}{h_{\mathfrak{g}}(x)} dt^2+\frac{x^2}{h_{\mathfrak{g}}(x)}(dx^\ast_{\mathfrak{g}})^2+x^2 (d\theta^2+\sin^2\theta d\varphi^2).
\end{equation}

Notice that the coordinate $(x,t)$ is singular at $x=x_{\text{h}}$. Thus, it is necessary to introduce  the  generalized Kruskal coordinates in the region  $x_{\text{h}}\leq x\leq x_{\text{h}}+\delta$ . Let us define \cite{Peltola:2008jx}
\begin{equation}
V_{\mathfrak{g}}:=t+x^\ast_{\mathfrak{g}},\quad U_{\mathfrak{g}}:=t-x^\ast_{\mathfrak{g}}.
\end{equation}
and
\begin{equation}
u_{\mathfrak{g}}:=\frac{1}{2}(e^{\xi_{\mathfrak{g}} V_{\mathfrak{g}}}+e^{-\xi_{\mathfrak{g}} U_{\mathfrak{g}}}),\quad  v_{\mathfrak{g}}:=\frac{1}{2}(e^{\xi_{\mathfrak{g}} V_{\mathfrak{g}}}-e^{-\xi_{\mathfrak{g}} U_{\mathfrak{g}}})
\end{equation}
with $\xi_{\mathfrak{g}}:=\left.\frac{1}{2}\partial_x(\frac{x^2}{h_{\mathfrak{g}}(x)})\right|_{x=x_{\text{h}}}$.
Using these coordinates, the spacetime metric reads  \cite{Peltola:2008jx}
\begin{equation}\label{metric3}
ds^2=\frac{x^2}{\xi_{\mathfrak{g}}^2h_{\mathfrak{g}}(x)}e^{-2\xi_{\mathfrak{g}} x^\ast_{\mathfrak{g}}}(-dv_{\mathfrak{g}}^2+du_{\mathfrak{g}}^2)+x^2 (d\theta^2+\sin^2\theta d\varphi^2).
\end{equation}
  It is easy to see that this metric is regular at $x=x_{\text{h}}$.
 Now, let us choose a point $u_{\mathfrak{g}}=v_{\mathfrak{g}}=0$ on the horizon  $x=x_{\text{h}}$, and define a new set of coordinates by 
\begin{equation}
T_{\mathfrak{g}}:=\lambda_{\mathfrak{g}} v_{\mathfrak{g}},\quad X_{\mathfrak{g}}:=\lambda_{\mathfrak{g}} u_{\mathfrak{g}}
\end{equation}
with $\lambda_{\mathfrak{g}}:=\lim_{x\to x_{\text{h}}}(\frac{\sqrt{x^2/h_{\mathfrak{g}}(x)}}{|\xi_{\mathfrak{g}}|e^{-\xi _{\mathfrak{g}}x^\ast_{\mathfrak{g}}}})$. Using these definitions, the spacetime metric on the given surface $u=v=0$ takes the formulation \cite{Peltola:2008jx,PADMANABHAN200549}
\begin{equation}
ds^2=-(dT_{\mathfrak{g}})^2+(dX_{\mathfrak{g}})^2+x^2 (d\theta^2+\sin^2\theta d\varphi^2).
\end{equation}
It is checked that a certain freely falling observer is momentarily at rest in this coordinate system $(T_{\mathfrak{g}},X_{\mathfrak{g}})$  at the surface $u_{\mathfrak{g}}=v_{\mathfrak{g}}=0$ \cite{Peltola:2008jx}.

Now, let us consider the  Klein-Gordan equation for the massless scalar field in the tortoise coordinate and the generalized Kruskal coordinate, respectively.
In the tortoise coordinate $x^\ast_{\mathfrak{g}}$, the Klein-Gordan equation for the massless scalar field $\phi(t,x^\ast_{\mathfrak{g}})=:\frac{R(t,x^\ast_{\mathfrak{g}})}{x}$ in the region  $x_{\text{h}}< x< x_{\text{h}}+\delta$ is given by
\begin{equation}\label{KG1}
\left(\frac{\partial^2}{\partial t^2}-\frac{\partial^2}{(\partial x^\ast_{\mathfrak{g}})^2}+V(x)\right) R(t,x^\ast_{\mathfrak{g}})=0,
\end{equation}
where the “potential” $V(x):=\frac{(\frac{x^2}{h_{\mathfrak{g}}(x)})\partial_x(\frac{x^2}{h_{\mathfrak{g}}(x)})}{x}$ satisfies
\begin{equation}
\lim_{x\to x_{\text{h}}}V(x)=0.
\end{equation}
By solving the Klein-Gordon equation  in the region  $x_{\text{h}}< x< x_{\text{h}}+\delta$ , one obtains the 
solutions 
\begin{equation}\label{solu1}
S_{k}=C\frac{e^{-\mathbf{i}\omega t+\mathbf{i}k x^\ast_{\mathfrak{g}}}}{x},\quad \omega=|k|
\end{equation}
which compose an orthonormal basis for $\phi$  in the region  $x_{\text{h}}< x< x_{\text{h}}+\delta$ . Moreover, we are interested in the behavior of the Klein-Gordan equation  in the coordinate $\check{x}^\ast_{\mathfrak{g}}$ given by the coordinate transformation $\check{x}^\ast_{\mathfrak{g}}=x^\ast_{\mathfrak{g}}+L-\delta^\ast_{\mathfrak{g}}$. It is direct to give the corresponding solution of the Klein-Gordan equation  in the coordinate $\check{x}^\ast_{\mathfrak{g}}$, which reads
\begin{equation}\label{solu100}
\check{S}_{k}=C\frac{e^{-\mathbf{i}\omega t+\mathbf{i}k \check{x}^\ast_{\mathfrak{g}}}}{x}
\end{equation}
 in the region  $x_{\text{h}}< x< x_{\text{h}}+\delta$ .
Then, the quantum scalar field $\hat{\phi}(t,x^\ast_{\mathfrak{g}})$  in the region  $x_{\text{h}}< x< x_{\text{h}}+\delta$  can be expanded as 
\begin{equation}
\hat{\phi}=\int dk (a^{\mathfrak{g}}_k\check{S}_{k}+a^{\mathfrak{g},\dagger}_k\check{S}^\ast_{k}),
\end{equation}
where $\check{S}^\ast_{k}$ is the complex conjugate of $\check{S}_{k}$.

In the generalized Kruskal coordinate in the region  $x_{\text{h}}\leq x<x_{\text{h}}+\delta$ , we can write down the Klein-Gordan equation for the massless scalar field $\phi(T_{\mathfrak{g}},X_{\mathfrak{g}})=:\frac{\tilde{R}(T_{\mathfrak{g}},X_{\mathfrak{g}})}{x}$ in a rest frame of the freely falling observer, which reads \cite{Peltola:2008jx}
\begin{equation}\label{KGK}
\left(\frac{\partial^2}{\partial T_{\mathfrak{g}}^2}-\frac{\partial^2}{\partial X_{\mathfrak{g}}^2}+\tilde{V}(x)\right) \tilde{R}(T_{\mathfrak{g}},X_{\mathfrak{g}})=0,
\end{equation}
where 
\begin{equation}
\tilde{V}(x):=\frac{1}{x}\frac{(\frac{x^2}{h_{\mathfrak{g}}(x)})\partial_x(\frac{x^2}{h_{\mathfrak{g}}(x)})}{\lambda_{\mathfrak{g}}^2 \xi_{\mathfrak{g}}^2}e^{-2\xi_{\mathfrak{g}} x^\ast_{\mathfrak{g}}}
\end{equation}
with
\begin{equation}
\lim_{x\to x_{\text{h}}}\tilde{V}(x)=\left.\frac{\partial_x(\frac{x^2}{h_{\mathfrak{g}}(x)})}{x}\right|_{x= x_{\text{h}}}.
\end{equation}
Consider the case that $x_{\text{h}}$ is large so that  $\tilde{V}(x)$ is small enough to be neglected. Then, by solving the Klein-Gordon equation \eqref{KGK} in the region  $x_{\text{h}}\leq x<x_{\text{h}}+\delta$ , one obtains orthonormal solutions 
\begin{equation}\label{solu2}
\tilde{S}_{k'}=C'\frac{e^{-\mathbf{i}\omega' T_{\mathfrak{g}}+\mathbf{i}k' X_{\mathfrak{g}}}}{x}.
\end{equation}
In addition, the quantum scalar field $\hat{\phi}$ in the region  $x_{\text{h}}< x<x_{\text{h}}+\delta$  can be expanded as 
\begin{equation}
\hat{\phi}=\int dk'(A^{\mathfrak{g}}_{k'}\tilde{S}_{k'}+A^{\mathfrak{g},\dagger}_{k'}\tilde{S}^\ast_{k'}),
\end{equation}
where $\tilde{S}^\ast_{k'}$ is the complex conjugate of $\tilde{S}_{k'}$.

Let us then consider the Bogolubov transformation between $(A^{\mathfrak{g}}_k, A^{\mathfrak{g},\dagger}_k)$ and  $(a^{\mathfrak{g}}_k, a^{\mathfrak{g},\dagger}_k)$.
Specifically, one has 
\begin{eqnarray}\label{AAdag}
A^{\mathfrak{g}}_{k'}=\int{dk}(\alpha_{kk'}a^{\mathfrak{g}}_{k}+\bar{\beta}_{kk'}a_{k}^{\mathfrak{g},\dagger}),\quad  A^{\mathfrak{g},\dagger}_{k'}=\int{dk}(\bar{\alpha}_{kk'}a^{\mathfrak{g},\dagger}_{k}+{\beta}_{kk'}a^{\mathfrak{g}}_{k}).
\end{eqnarray}
Recall the solution \eqref{solu2} of the Klein-Gordan equation in the   generalized Kruskal coordinate, one can know that the Bogolubov coefficients $\alpha_{ k'k }$ and $\beta_{k'k }$ are determined by the expansion
\begin{equation}\label{Btrans111}
e^{-\mathbf{i}k\check{U}_{\mathfrak{g}}}=\int{dk'}(\alpha_{kk' }e^{-\mathbf{i}k'\tilde{U}_{\mathfrak{g}}}+\beta_{kk' }e^{\mathbf{i}k'\tilde{U}_{\mathfrak{g}}}),
\end{equation}
where $\tilde{U}_{\mathfrak{g}}:=T_{\mathfrak{g}}-X_{\mathfrak{g}}$ and
\begin{equation}
\check{U}_{\mathfrak{g}}:=t-\check{x}^\ast_{\mathfrak{g}}=-\xi_{\mathfrak{g}}^{-1}\ln(-\tilde{U}_{\mathfrak{g}})+\xi_{\mathfrak{g}}^{-1}\ln\lambda_{\mathfrak{g}}+L-\delta^\ast_{\mathfrak{g}}.
\end{equation}  
It is verified that the Bogolubov coefficients $\alpha_{kk' }$ and $\beta_{kk' }$ satisfy \cite{Peltola:2008jx}
\begin{equation}
|\alpha_{kk' }|=e^{\pi\xi_{\mathfrak{g}}^{-1}\omega_k}|\beta_{kk' }|.
\end{equation}
Then, one can define the local Hartle-Hawking vacuum  $|0_{\text{HH}} \rangle_{\mathfrak{g}}$ in the region $x_{\text{h}}\leq x<x_{\text{h}}+\delta$ by  
\begin{eqnarray}
A^{\mathfrak{g}}_k|0_{\text{HH}}\rangle_{\mathfrak{g}}=0,\quad \forall k.
\end{eqnarray}
This leads to \cite{PADMANABHAN200549 }
\begin{equation}
|0_{\text{HH}} \rangle_{\mathfrak{g}}=\prod_{{k} }\left({(1-e^{-2\pi\xi_{\mathfrak{g}}^{-1}\omega_{{k} }})^{1/2}} \sum^{\infty}_{n_{k}=0 }e^{-\pi n_{k} \omega_{k}/\xi_{\mathfrak{g}}}\frac{(a^{\mathfrak{g},\dagger}_{k})^{n_{k}}}{(\mathcal{N})^{n_k}\sqrt{n_k!}}\right)|0_{\text{B}}\rangle_{\mathfrak{g}},
\end{equation}
where $n_{k}=0,1,2,3,...$ is the particle number on the mode $k$, $\mathcal{N}:=\frac{1}{\sqrt{2\pi\hbar \delta(k-k')|_{k=k'}}}$ is a normalization constant. Equivalently,  $|0_{\text{HH}} \rangle_{\mathfrak{g}}$ can be rewritten as
\begin{equation}\label{0Kk222}
|0_{\text{HH}} \rangle_{\mathfrak{g}}= \sum_{\vec{n} }\left(\prod_{{k} }{(1-e^{-2\pi\xi_{\mathfrak{g}}^{-1}\omega_{{k} }})^{1/2}}e^{-\pi n_{k} \omega_{k}/\xi_{\mathfrak{g}}}\right)|\vec{k}_{\vec{n}}\rangle_{\mathfrak{g}},
\end{equation}
where 
\begin{equation}
|\vec{k}_{\vec{n}}\rangle_{\mathfrak{g}}:=\left(\prod_{{k} }\frac{(a^{\mathfrak{g},\dagger}_{k})^{n_{k}}}{(\mathcal{N})^{n_k}\sqrt{n_k!}}\right)|0_{\text{B}}\rangle_{\mathfrak{g}},
\end{equation}
and 
\begin{equation}
\vec{n}=(n_{k_1},n_{k_2},n_{k_3},...)
\end{equation}
with the sequence $k_1<k_2<k_3<...$ composed of all possible values in the spectrum of $k$.

\end{document}